**Highly conserved sequence-specific double-stranded DNA binding networks contributing to divergent genomic evolution of human and chimpanzee brain development.**

Gennadi V. Glinsky[1]

[1] Institute of Engineering in Medicine
University of California, San Diego
9500 Gilman Dr. MC 0435
La Jolla, CA 92093-0435, USA
Correspondence:
genlighttech@gmail.com

**Abstract**

Emergence during mammalian evolution of concordant and divergent traits of genomic regulatory networks encompassing ubiquitous, qualitatively nearly identical yet quantitatively distinct arrays of sequences of transcription factor binding sites (TFBS) for 716 proteins is reported. A vast majority of TFs (770 of 716; 98%) comprising protein constituents of these networks appear to share common Gene Ontology (GO) features of sequence-specific double-stranded DNA binding (GO: 1990837). Genome-wide and individual chromosome-level analyses of 17,935 ATAC-seq-defined brain development regulatory regions (BDRRs) revealed nearly universal representations of TFBS for TF-constituents of these networks, TFBS densities of which appear consistently higher within thousands BDRRs of Modern Humans compare to Chimpanzee. Transposable elements (TE), including LTR/HERV, SINE/Alu, SVA, and LINE families, appear to harbor and spread genome-wide consensus regulatory nodes of identified herein highly conserved sequence-specific double-stranded DNA binding networks, selections of TFBS panels of which manifest individual chromosome-specific profiles and species-specific divergence patterns. Collectively, observations reported in this contribution highlight a previously unrecognized essential function of human genomic DNA sequences encoded by TE in providing genome-wide regulatory seed templates of highly conserved sequence-specific double-stranded DNA binding networks likely contributing to continuing divergent genomic evolution of human and chimpanzee brain development.

**Introduction**

DNA sequences derived from transposable elements (TE) constitute ~50% of human genome, contributing to a multitude of structural features and regulatory functions at different levels of genomic organization (Elbarbary et al., 2016; Sundaram and Wysocka, 2020). During evolution, concurrently with TE colonization of primates' genomes, a highly sophisticated defense system co-evolved designed to restrict uncontrolled TE expansion and diminish potentially deleterious effects of TE insertions on genome integrity, while providing TE family sequence-specific co-option mechanisms integrating TE-derived sequences into cell type- and tissue-specific genomic regulatory networks (Jacobs et al., 2014; Pontis et al., 2019). While potentially destructive effects of TE on genome integrity represent their ubiquitous biological feature defined by the very nature of transpositionally-competent TE, it is unknown whether TE-derived sequences may possess similarly ubiquitous non-deleterious functions, perhaps, contributing to evolutionary fitness of the host.

TE-derived sequences are often integrated into species' genomics regulatory networks. In mammalian genomes, TE-derived sequences rewired the core regulatory circuitry of embryonic stem cells (Kunarso et al., 2010); they intrinsically activated in human preimplantation embryos (Grow et al., 2015) by operating as long-range enhancers (Fuentes et al, 2018) and providing transcription factor binding sites (TFBS) for TP53 (Wang et al., 2007), STAT1 (Schmid et al., 2010), CTCF (Schmidt et al., 2012), a set of transcription factors (TFs) defined as master pluripotency regulators (Kunarso et al., Schmidt et al., 2012), including candidate human-specific TFBS for POU5F1 (OCT4), NANOG, SOX2, and CTCF (Glinsky, 2015; Glinsky et al., 2018). Overall, chromatin immunoprecipitation and sequencing (Chip-seq) experiments demonstrated that TE-derived loci harbor thousands of TFBS and presumably exerting global regulatory effects on gene expression in various pathophysiological conditions. However, distribution of TFBS are non-uniform and highly variable within specific TE subfamilies that evolutionary emerged from identical or highly similar sequences. TE sequence-dependent distribution of TFBS is likely influenced by genetic drift and other mutational processes as well as cell type-specific epigenetic contexts affecting chromatin states of genomic regions harboring TE insertions. Based on these considerations, it was reasonable to assume that Chip-seq experiments could capture only a fraction of TFBS repertoire encoded by the ancestral TE loci that may have existed at the time of the TE insertion and expansion in a host genome. Therefore, it was of interest to catalogue all TFs that have potential TFBS located within genomic loci encoded by a specific TE subfamily, which should facilitate the comparative analyses of TFBS encoded by distinct TE families to infer their concordant, non-overlapping, and discordant regulatory functions.

**Results**

**Highly conserved sequence-specific double-stranded DNA binding genomic regulatory networks (GRNs) encoded by distinct families of human embryo regulatory LTRs.**

To catalogue all TFs that have potential TFBS located within genomic loci encoded by a specific TE subfamily of human embryo regulatory LTRs (Glinsky, 2022; 2024), the Jaspar algorithm was employed to identify all TFBS located within 606 LTR5_Hs loci previously characterized as functionally active regulatory elements in human cells (Fuentes et al, 2018; Glinsky, 2022). The LTR5_Hs loci are likely to manifest the relatively minor divergence of TFBS profiles because LTR5_Hs/HERVK successfully infected and colonized primates' germline most recently compared to other HERVs with documented regulatory functions in human embryogenesis (Glinsky, 2022; 2024). The Jaspar algorithm output was processed to identify at a single-nucleotide resolution all TFs having predicted TFBS within LTR5_Hs loci and the number of TFBS for each TF was calculated (Supplementary Table S1). Overall, LTR5_Hs loci harbor 771 distinct TFBS for 716 individual TFs, including 65 paired TFBS features for 68 TFs comprising the immediately adjacent TFBS for two different TFs. Since individual LTR5_Hs loci harbor subsets of a composite set of 771 TFBS, it was designated a composite ancestral set of TFBS encoded by LTR5_Hs loci. Gene Set Enrichment Analyses (GSEA) employing the ENRICHR bioinformatics platform documented the exceedingly broad engagement of the LTR5_Hs loci-residing TFs in manifestations of physiological phenotypes and pathological conditions of Homo sapiens (Table 1; Supplementary Table S1). As expected, GSEA employing GO Molecular Function 2023 database identified among most significantly enriched categories 556 genes of Cis-Regulatory Region Sequence-Specific DNA Binding (GO: 0000987) category; 581 genes of RNA Polymerase II Cis-Regulatory Region Sequence-Specific DNA Binding (GO: 0000978) category; 609 genes of RNA Polymerase II Transcription Regulatory Region

Sequence-Specific DNA Binding (GO:0000977) category. Notably, 700 of 716 (98%) of TFs having TFBS within LTR5_Hs loci represent constituents of the single Gene Ontology classification category designated the Sequence-Specific Double-Stranded DNA Binding (GO: 1990837). These observations indicate that identified herein set of LTR5_Hs-residing TFs may constitute a sequence-specific double-stranded DNA binding pathway exerting exceedingly broad and predominantly regulatory impacts on phenotypes of Modern Humans.

Quantitative features of the ancestral LTR5_Hs set of TFBS were documented by calculating the numbers of TFBS for each of the 771 distinct TFBS and reporting the computed values as numbers of TFBS per LTR5_Hs locus (defined as TFBS frequency) and the estimated TFBS density normalized to 1 Kb of the locus length (defined as TFBS density; Supplementary Table S1). It has been observed that 533 of 771 distinct TFBS (69%) have at least one TFBS per LTR5_Hs locus and TFBS density more than 1 binding site per 1 Kb of regulatory DNA, while top 100 TFBS have TFBS frequency of more than 8.7 binding sites per LTR5_Hs locus and TFBS density more than 9.9 binding sites per 1 kb of regulatory DNA. These observations suggest that a significant fraction of the LTR5_Hs ancestral set of TFBS has been preserved in LTR5_Hs loci residing in the human genome.

Analysis of TFBS density values indicate that individual LTR5_Hs loci harbor TFBS for numerous TFs (Supplementary Table S1). This feature appears similar to experimentally defined 3583 regulatory loci in the mouse ESC which are documented to have TFBS for multiple TFs (Chen et al., 2008). It was of interest to identify all potential TFBS residing within 3583 mouse ESC multi-TFs-binding regulatory loci and compared them to the ancestral LR5_Hs set of 771 TFBS. These analyses demonstrated that 3583 mouse ESC multi-TFs-binding loci harbor 776 distinct TFBS (Supplementary Table S2). Notably, all 771 distinct TFBS comprising the ancestral set of LTR5_Hs-encoded TFBS were identified among TFBS residing within mouse ESC regulatory sequences binding multiple TFs. These observations suggest that identified herein ancestral set of 771 TFBS encoded by LTR5_Hs elements may reflect the presence of evolutionary conserved genomic regulatory network (GRN) operating during mammalian embryogenesis.

It was of interest to characterize the spectrum of TFBS encoded by other LTR families known to play a regulatory role during human embryogenesis and collectively defined as human embryo regulatory LTRs (Glinsky, 2022; 2024). To this end, all potential TFBS encoded by LTR7; MLT2A1; and MLT2A2 loci were identified employing the Jaspar algorithm and corresponding quantitative metrics were computed for each set of TFBS (Supplementary Table S3). Intriguingly, despite clearly discernable marked divergence of DNA sequences, all analyzed herein regulatory LTR elements appear to harbor nearly identical sets of TFBS (Figure 1; Supplementary Table S3). Since introductions into primates' germline of different families of analyzed herein regulatory LTRs were separated by millions of years, these observations are consistent with the hypothesis of evolutionary conservation of identified in this contribution GRN. In agreement with this concept, correlation analyses of TFBS density profiles (Figure 1) revealed highly concordant patterns of TFBS densities between mouse ESC multi-TFs-binding loci and distinct families of human embryo regulatory LTRs, namely MLTA1A2 (r = 0.948); MLT2A1 (r = 0.921); LTR7 (r = 0.900); and LTR5_Hs (r = 0.834). Interestingly, networks of TFBS encoded by evolutionary ancient LTRs (MLT2A2 and MLT2A1) appear more closely related to the TFBS profile of mouse ESC multi-TFs-binding regulatory loci compared to TFBS networks encoded by human embryo regulatory LTRs introduced into primate germlines relatively recently (LTR7 and LTR5_Hs).

Despite overall similarities TFBS density profiles (Figure 1), human embryo regulatory LTRs manifest notable quantitative differences of the TFBS numbers for specific TFs (Figure 2; Supplementary Figure S1; Supplementary Table S4). Interestingly, LTR5_Hs loci appear to exhibit the largest divergence of the TFBS density profiles when comparisons were made to either mouse ESC multi-TFs-binding loci or other families of human embryo regulatory LTRs, including MLT2A1; MLT2A2; and LTR7 sequences (Figure 2; Supplementary Figure S1; Supplementary Table S4). For example among LTR5_Hs TFs with top 100 TFBS density scores, 19 TFs have at least 50% higher TFBS density compared to mouse ESC multi-TFs-binding regulatory loci and no TFs have lower TFBS density at this threshold. In contrast, within LTR7 loci there are 9 TFs manifesting at least 50% TFBS density gain and 4 TFs exhibiting TFBS density loss compared to mouse ESC multi-TFs-binding regulatory loci. Similarly, within MLT2A1 loci there are 7 TFs manifesting gain and 8 TFs exhibiting loss of at least 50% TFBS density compared to mouse ESC multi-TFs-binding regulatory loci. There were no TFs with TFBS density gain of at least 50% and 4 TFs with TFBS density loss within MLT2A2 loci. Direct comparisons of TFBS density scores between different families of human embryo regulatory LTRs reveals that among LTR5_Hs TFs with top 100 TFBS density scores, there are 23; 33; and 36 TFs having at least 50% higher TFBS density scores within LTR5_Hs loci compared to LTR7; MLT2A2; and MLT2A1 loci, respectively. In contrast, there are 1; 1; and 7 TFs having at least 50% higher TFBS density scores within LTR7; MLT2A2; and MLT2A1 loci compared to LTR5_Hs loci, respectively. Comparative analyses of TFBS density scores between LTR5_Hs and LTR7 loci among TFs with top 533 TFBS density scores (see above), demonstrate that there are 159 TFs and 55 TFs manifesting at least 50% higher TFBS density scores within LTR5_Hs versus LTR7 loci and LTR7 versus LTR5_Hs loci, respectively. At the threshold of at least 100% gains of TFBS density scores, there are 107 TFs having higher TFBS density scores within LTR5_Hs loci versus LTR7 loci, whereas there are no TFs harboring higher TFBS density scores within LTR7 loci versus LTR5_Hs loci (Supplementary Table S1). These findings suggest that LT5_Hs loci may have made a significant impact on the divergence of GRNs governed by the human embryo regulatory LTRs by providing regulatory loci with increased numbers of TFBS for numerous TFs (Figure 2; Supplementary Figure S1; Supplementary Tables S1 and S4). GSEA of TF-coding genes manifesting most significant increases of TFBS densities within LTR5_Hs loci compared to either mouse ESC multi-TFs-binding loci and/or other human embryo regulatory LTRs demonstrated that these TFs contribute to development of preimplantation embryogenesis phenotypic traits (Embryonic stem cells; Trophoblast stem cells; Blastocyst; Germ cells; Ectoderm); as well as multiple types of cells and tissues of central nervous system, including Neural stem/precursor cells; Neural crest; Neuroblast; Neural tube; Motor neurons; Oligodendrocyte precursor/progenitor cells; Fetal brain; Cerebellum; Prefrontal cortex; Superior frontal gyrus; Peripheral nerve (Supplementary Figure S1; Supplementary Tables S1 and S4). These findings are consistent with the previously reported observations based on gene ontology-guided proximity placement analyses of GRNs governed by human embryo regulatory LTRs, which implicated LTR-associated GRNs in regulation of development and functions of primates' central nervous system (Glinsky, 2022; 2024).

**Comparative analyses of highly conserved sequence-specific double-stranded DNA binding GRNs within ATAC-seq-defined regulatory regions of human and chimpanzee brain development.**

It was of interest to extend further this line of inquiry by determining whether reported herein sets of TFs and TFBS constituting the Sequence-Specific Double-Stranded DNA Binding GRNs associated with human embryo regulatory LTRs could be identified within genomic regulatory loci known to contribute to regulation of human and chimpanzee brain development. To this end, 17,935 genomic regulatory regions defined in the organoid single-cell genomic atlas of human and chimpanzee brain development (Kanton et al, 2019) were analyzed to catalogue all TFs having putative TFBS within these regions (Table 2). Kanton et al (2019) reported 8099 human-specific and 9836 chimpanzee-specific brain development regulatory regions (BDRRs), which were identified employing DNA sequence unbiased open chromatin accessibility screening method termed the Assay for Transposase-Accessible Chromatin with high-throughput sequencing (ATAC-seq; Buenrostro et al, 2013; 2015). Genome-wide and individual chromosome level analyses of these regions revealed that TFs and corresponding TFBS of the Sequence-Specific Double-Stranded DNA Binding pathway comprise the intrinsic elements of the ATAC-seq-defined human and chimpanzee BDRRs (Table 2). Notably, quantitative characteristics of the Sequence-Specific Double-Stranded DNA Binding networks appear significantly distinct between human and chimpanzee ATAC-seq-defined regulatory regions of brain development. While sets of individual TFs and corresponding TFBS independently identified in both species are indistinguishable, the TFBS frequencies (numbers of TFBS per ATAC-seq-defined regulatory locus) and the TFBS densities (TFBS numbers per 1 kb of ATAC-seq-defined regulatory locus) are consistently higher in Modern Humans compared to Chimpanzee BDRRs (Table 2). These finding suggest that reported distinct quantitative characteristics of the Sequence-Specific Double-Stranded DNA Binding pathway within ATAC-seq-defined brain development regulatory loci may have contributed to the previously observed different trajectories of Modern Humans and Chimpanzee brain development (Kanton et al., 2019).

Next, quantitative characteristics of gains and/or losses for individual LTR5_Hs loci-associated TFs having top 100 TFBS density scores (see above) were calculated in 17,935 ATAC-seq-defined BDRRs of human and chimpanzee, corresponding quantitative metrics were estimated for individual species' chromosomes, and multiple comparative analyses were carried out. Comparisons of TBS density profiles within BDRRs of the individual chromosomes of human and chimpanzee genomes that were acquired as a result of TFBS density gains or losses during mammalian evolution (Figure 3) identified five chromosomes manifesting highly concordant changes of TFBS densities, namely chr19 (r = 0.99); chr22 (r = 0.98); chr4 (r = 0.94); chr17 (r = 0.93); chr13 (r = 0.90); while eight chromosome exhibit weak or no concordance of TFBS density changes, namely chr21 (r = -0.12); chr14 (r = 0.12); chrX (r = 0.31); chr9 (r = 0.35); chr7 (r = 0.37); chr11 (r = 0.40); chr15 (0.43); chr12 (r = 0.47). Examples of correlation plots for two chromosomes manifesting most concordant (chr19 and chr22) and discordant (chr21 and chr14) profiles of TFBS density changes are shown in the Figure 3. Remaining 10 chromosomes appear to have moderate concordance levels of TFBS density profile changes within BDRRs of human and chimpanzee (Figure 3), suggesting largely divergent patterns of evolutionary TFBS density gains and losses of two species.

To explore further the hypothesis of the divergent evolution of TFBS density gains within BDRRs of human and chimpanzee, TFBS density gains and losses compared to mouse ESC multi-TFs-binding loci were computed for each chromosome and numbers of events manifesting TFBS density changes of at least 50% were plotted for visualization (Figure 4). Results of these analyses documented clearly discernable distinct patterns of TFBS density gains and losses acquired by humans and chimpanzee during mammalian evolution (Figure 4). Interestingly, humans appear to manifest predominantly gains of TFBS density within BDRRs, while chimpanzee seem to exhibit the prevalent losses within regulatory regions of brain development at several chromosomes (Figure 4). These observations were corroborated by direct comparisons of TFBS density values within BDRRs of human versus chimpanzee, which were calculated for each individual TFBS, recorded, and reported for each chromosome as the numbers of events manifesting most significant species-specific gains of TFBS densities (Figure 4).

Computation of gains and losses for individual TFs of the Sequence-Specific Double-Stranded DNA Binding networks in 17,935 brain development genomic regulatory regions and recording of corresponding quantitative metrics of TFBS density changes for all individual chromosomes of human and chimpanzee genomes facilitated both interspecies and within species comparative genome-wide chromosome-level analyses.

To this end, genome-wide chromosome-level pairwise correlation matrices of TFBS density changes acquired during mammalian evolution within human (Table 3) and chimpanzee (Table 4) BDRRs defined by the ATAC-seq analysis were developed and analyzed for within specie's genome concordance and divergence patterns. Using numerical values of correlation coefficients reported in Tables 3 and 4, the mean values of correlation coefficients were calculated for each chromosome of human and chimpanzee genomes and corresponding divergence scores were quantified by subtracting the mean values from a perfect correlation coefficient value of 1.0. Additionally, for each individual human and chimpanzee chromosome coefficient of variations were estimated as the ratio of standard deviation to the mean value expressed as a percentage. Corresponding numerical values were plotted for visualization and reported in the Figure 5. In human genome, 19 chromosomes appear to exhibit similar values of divergence scores ranging from 0.208 (chr10) to 0.325 (chr20). In contrast, seemingly higher values of divergence scores were documented for 4 chromosomes, namely 0.553 (chr16); 0.831 (chr17); 0.960 (chr22); and 1.047 (chr19). These findings were corroborated by the results of the analysis of variation coefficients which identified chr17; chr19; and chr22 as chromosomes having largest values of coefficients of variation among human chromosomes (Figure 5). In chimpanzee genome, values of divergence scores appear to manifest a somewhat broader degree of variability ranging from 0.194 (chr7) to 0.472 (chr6), while 2 chromosomes had seemingly higher divergence score values of 0.646 (chr13) and 0.757 (chr4). Consistent with these observations, results of the analysis of variation coefficients identified chr13 and chr4 as chromosomes associated with largest values of coefficients of variation in chimpanzee genome (Figure 5).

Therefore, analyses of patterns of divergence and concordance of chromosome-level TFBS density changes acquired by Modern Humans and Chimpanzee during mammalian evolution identified chr17; chr19; and chr22 as chromosomes exhibiting most divergent profiles of TFBS density changes in human genome, while chr4 and chr13 appear to manifest most divergent patterns of TFBS density changes in chimpanzee genome (Figure 5). Notably, chromosomes that appear to manifest most divergent patterns of TFBS density changes within human (chr17; chr19; and chr22) or chimpanzee (chr4; chr13) genomes are the same chromosomes that have most similar profiles of TFBS density gains/losses in brain development regulatory regions of human and chimpanzee (Figure 4). However, most divergent chromosomes within chimpanzee genome, namely chr4 and chr13, manifest variation coefficients closely related to 18 other chromosomes within human genome (Figure 5). Conversely, chr17; chr19; and chr22 that are most divergent within human genome have coefficients of variation closely related to 18 other chromosomes of chimpanzee genome (Figure 5).

**Distinct and common association patterns of TFBS density changes within BDRRs of human and chimpanzee revealed by genome-wide chromosome level alignment analyses with signatures of TFBS density changes of different families of human embryo regulatory LTRs.**

Establishment of genome-wide chromosome-level quantitative profiles of TFBS density changes within ATAC-seq-defined BDRRs of human and chimpanzee (Tables 2 – 4; Figures 3 – 5) prompted investigation of association patterns of TFBS density changes within BDRRs and signatures of TFBS density changes of 4 different families of human embryo regulatory LTRs (Figure 6). Visualization of the results of chromosome-level alignments of the corresponding profiles of TFBS density changes revealed clearly discernable common and distinct patterns of associations of TFBS density changes acquired during mammalian evolution within human and chimpanzee BDRRs and within DNA sequences encoded by different families of human embryo regulatory LTRs, namely MLT2A1 (Figure 5A); MLT2A2 (Figure 6C); LTR7 (Figure 6D); and LR5_Hs (Figure 6B). Notably, the concordant and discordant association patterns of TFBS density changes were observed in interspecies comparisons of TFBS density changes as well as in analyses of within-specie profiles of TFBS density changes of BDRRs aligned to signatures of TFBS density changes of different families of regulatory LTRs (Figure 6). Alignments to TFBS density changes profiles of BDRRs and the MLT2A1 and LTR5_Hs generated larger values of correlation coefficients compared to the MLT2A1 and LTR7 alignments, while highly concordant alignment patterns were observed for MLT2A1 and MLTA2 analyses as well as for LTR5_Hs and LTR7 analyses. These trends were observed in interspecies and within individual specie comparisons. Comparisons of within an individual specie genome alignments revealed striking negative correlations of the association patterns generated by the alignments of BDRRs TFBS density changes profiles to TFBS density changes profiles of the MLT2A1 versus LTR5_Hs regulatory LTRs (Figure 6), which were documented in analyses of either chimpanzee or human BDRRs. However, BDRRs residing only on 4 human chromosomes, namely chr19; chr22; chr17; and chr6, manifested positive correlation coefficients of TFBS density changes profiles alignments to the LTR5_Hs in contrast to BDRRs residing on 12 chimpanzee chromosomes. Conversely, BDRRs housed on 19 human chromosomes had positive correlation coefficients of TFBS density changes profiles alignments to the MLT2A1 loci in contrast to BDRRs housed on 10 chimpanzee chromosomes (Figure 6). These findings were corroborated by the results of the interspecies comparisons of the alignments of TFBS density changes profiles of human and chimpanzee BDRRs the profiles of TFBS density changes of either MLT2A1 or LTR5_Hs (Figure 6). Combining these alignments into one plot has resulted in the visual depiction of common and distinct patterns of TFBS density changes within BDRRs residing on different chromosomes of human and chimpanzee genomes (Figure 6). Overall, these observations are in agreement with the hypothesis that divergence of TFBS density changes within BDRRs residing on different chromosomes of human and chimpanzee genomes occurs along distinct alignment patterns to profiles of TFBS density changes of different families of human embryo regulatory LTRs, namely MLT2A1 and LTR5_Hs. This model was further corroborated by results of the analyses aggregating the chromosome-level observations into a simplified genome-wide data set highlighting panels of TFs manifesting either common or divergent patterns of TFBS density changes acquired during primate evolution within BDRRs of human and chimpanzee (Figure 6; Supplementary Table S5). GSEA of TF-coding genes manifesting divergent profiles of TFBS within BDRRs of human and chimpanzee underscore their exceedingly broad developmental and pathophysiological impacts on phenotypic traits of Modern Humans (Supplementary Table S5), including key constituents of central nervous system development and functions.

**Potential impacts of currently active LINE and SVA transposons in shaping the continuing divergent genomic evolution of BDRRs of Modern Humans and Chimpanzee.**

Observations reported in this contribution are congruent with recent findings implicating TE-encoded regulatory sequences derived from multiple TE families in development of human and chimpanzee hippocampal intermediate progenitor cells (Patoori et al., 2022). It was of interest to investigate whether the TF-constituents of reported herein sequence-specific double-stranded DNA binding networks are engaged in TE-governed hippocampal neurogenesis regulatory pathways, which were discovered by Patoori et al. (2022) utilizing a model of differentiation of human and chimpanzee induced pluripotent stem cells into TBR2 (or EOMES)-positive hippocampal intermediate progenitor cells. Results of these analyses (Figure 7; Supplementary Table S6) documented ubiquitous presence within TE-constituents of regulatory pathways of hippocampal intermediate progenitors of qualitatively nearly identical arrays of TFBS for TFs constituting protein components of reported herein sequence-specific double-stranded DNA binding networks. These observations indicate that one of universal features of multiple families of TEs, including LTR/HERV, SINE/Alu, SVA, and LINE families, is their intrinsic propensity to harbor and spread genome-wide consensus regulatory nodes of identified herein highly conserved sequence-specific double-stranded DNA binding networks, selections of TFBS panels of which manifest individual chromosome-specific profiles and species-specific divergence patterns. Consistent with this hypothesis, it has been observed that TE subfamilies that became more divergent from consensus TFBS patterns due to mutational processes are less likely to be represented within differentially-accessible (DA) ATAC-seq-defined regulatory loci of hippocampal intermediate progenitors' development (Figure 7), while DA ATAC-regions intersecting larger numbers of SINE/Alu loci appear to harbor TFBS for more TF-constituents of highly conserved sequence-specific double-stranded DNA binding networks (Figure 7). Notably, different families of TE-constituents of regulatory pathways of hippocampal intermediate progenitors' development manifest different degrees of conservation of arrays of TFBS for a consensus panel of TF-constituents of highly conserved sequence-specific double-stranded DNA binding networks. SINE/Alu subfamily members seem to exhibit the highest diversity and least apparent conservation profiles reaching the maximum divergence of 63.4%. In contrast, LINE subfamily members manifest the maximum divergence of only 14.56%, while the maximum divergence observed for LTR subfamily members was 36.21% (Figure 7).

Intriguingly, TE families that are currently active as retrotransposons in human genome, namely LINE and SVA transposable elements, harbor arrays of TFBS for essentially all TF-constituents of highly conserved sequence-specific double-stranded DNA binding networks (Figure 7; Supplementary Table S6). Therefore, it was of interest to investigate association patterns of TFBS density changes within human and chimpanzee BDRRs and signatures of TFBS density changes acquired during mammalian evolution by SVA and LINE retrotransposons and compare the results with similar analyses carried out for human embryo egulatory LTRs (Figure 6). Visualization of the results of chromosome-level alignments of the corresponding profiles of TFBS density changes revealed clearly discernable common and distinct patterns of associations of TFBS density changes acquired during mammalian evolution within human and chimpanzee BDRRs and within DNA sequences encoded by SVA and L1PA6 loci (Figure 8). The concordant and discordant patterns of association of TFBS density changes were observed in interspecies comparisons of TFBS density changes as well as in analyses of within-specie profiles of TFBS density changes of BDRRs aligned to signatures of TFBS density changes of SVA and L1PA6 retrotransposons (Figure 8). Alignments to TRBS density changes profiles of BDRRs and the SVA and L1PA6 generated larger values of correlation coefficients compared to the MLT2A2 and LTR5_Hs alignments (Figure 6 and 8), while highly concordant alignment patterns were observed for MLT2A1 and L1PA6 analyses as well as for LTR5_Hs and SVA analyses. These trends were observed in interspecies and within individual specie comparisons. Comparisons of within an individual specie genome alignments revealed striking negative correlations of the association patterns of the alignments of BDRRs TFBS density changes profiles to TFBS density changes profiles of the L1PA6 versus LTR5_Hs and the L1PA6 versus SVA, in contrast to highly positive correlations of TFBS density changes profiles of the L1PA6 versus MLT2A1 (Figure 8). These patterns of associations were consistently observed in analyses of either chimpanzee or human BDRRs.

BDRRs housed on 19 human chromosomes had positive correlation coefficients of TFBS density changes profiles alignments to the L1PA6 loci in contrast to BDRRs housed on 6 chimpanzee chromosomes, including chr4 and chr13 (Figure 8). In contrast, BDRRs residing only on 6 human chromosomes, including chr19; chr22; chr17; and chr6, manifested positive correlation coefficients of TFBS density changes profiles alignments to the SVA in contrast to BDRRs residing on 17 chimpanzee chromosomes.

These findings were corroborated by the results of the interspecies comparisons of the alignments of TFBS density changes profiles of human and chimpanzee BDRRs the profiles of TFBS density changes of either L1PA6 or SVA (Figure 6). Combining these two alignments into a single plot has resulted in the visual depiction of common and distinct patterns of TFBS density changes within BDRRs residing on different chromosomes of human and chimpanzee genomes (Figure 8). These observations are in agreement with the hypothesis that divergence of TFBS density changes within BDRRs residing on different chromosomes of human and chimpanzee genomes occurs along distinct alignment patterns to profiles of TFBS density changes of retrotransposons that are currently active in human genome, namely SVA and LINE families.

**Discussion**

In this contribution, the emergence during mammalian evolution of genomic regulatory networks (GRNs) encompassing ubiquitous, qualitatively nearly identical and quantitatively markedly distinct arrays of sequences of TFBS for 716 proteins is reported. A vast majority of TFs (770 of 716; 98%) comprising protein constituents of these networks appear to share common Gene Ontology (GO) features of sequence-specific double-stranded DNA binding (GO: 1990837). Among most significantly enriched categories, GSEA employing GO Molecular Function 2023 database identified 556 genes assigned to Cis-Regulatory Region Sequence-Specific DNA Binding (GO: 0000987) category; 581 genes of RNA Polymerase II Cis-Regulatory Region Sequence-Specific DNA Binding (GO: 0000978) category; and 609 genes of RNA Polymerase II Transcription Regulatory Region Sequence-Specific DNA Binding (GO:0000977) category. Assignments of essentially all TFs constituents of these networks to GO functional categories of sequence-specific double-stranded DNA binding, cis-regulatory region sequence-specific DNA binding, RNA Polymerase II cis-regulatory and transcription regulatory region sequence-specific DNA binding strongly imply their structural-functional engagements into assembly and activities of heterochromatin and euchromatin multiprotein-DNA complexes. To date, ubiquitous, qualitatively nearly identical and quantitatively markedly distinct representations of sequence-specific TFBS arrays of these networks have been observed within genomic regulatory loci encoded by all analyzed TE families, including TE families coopted into GRNs contributing to development and functions of central nervous system. TE families, including LTR/HERV, SINE/Alu, SVA, and LINE subfamilies, appear to harbor and spread genome-wide consensus regulatory nodes of identified herein highly conserved GRNs, selections within which of TFBS panels manifest individual chromosome-specific profiles and species-specific divergence patterns. Markedly distinct quantitative characteristics of these networks, in particular, changes of TFBS densities, have been inferred from genome-wide chromosome-level analyses of BDRRs of Modern Humans and Chimpanzee, suggesting that species-specific differences of the activities of these networks may have contributed to continuing divergent genomic evolutions of brain development of humans and non-human primates. Reported in this contribution results of chromosome-level analyses of quantitative metrics of GRNs emanating from sequence-specific double-stranded DNA binding of ~700 proteins may achieve a marked functional diversity by operating in chromosome territory-specific patterns. Observed conservation of these GRNs beyond the boundaries of confidently mapped TE-derived regulatory loci suggest that considerations of contributions of TEs to creation of mammalian genomic DNA could be extended to more than currently estimated ~50% of genomes.

**Methods**

*Data source and analytical protocols*

Solely publicly available datasets and resources were used in this contribution. Initial analyses were focused on human embryo regulatory LTR loci (see Introduction) that were identified as highly-conserved pan-primate regulatory sequences because they have been present in genomes of primate species for at least ~15 MYA. Four distinct LTR families meeting these criteria, namely MLT2A1 (2416 loci), MLT2A2 (3069 loci), LTR7 (3354 loci), and LTR5_Hs (606 loci), were analyzed. A total of 9445 fixed non-polymorphic sequences of human embryo regulatory LTR elements residing in genomes of Modern Humans (hg38 human reference genome database) were retrieved as described in recent studies (Hashimoto et al., 2021; Carter et al., 2022; Glinsky, 2022; 2024) and the number of highly conserved orthologous loci in genomes of sixteen non-human primates (NHP) were determined exactly as previously reported (Glinsky, 2022; 2024). Briefly, fixed non-polymorphic regulatory LTR loci residing in the human genome (hg38 human reference genome database) has been considered highly conserved in the genome of NHP only if the following two requirements are met:

(1) During the direct LiftOver test (https://genome.ucsc.edu/cgi-bin/hgLiftOver ), the human LTR sequence has been mapped in the NHP genome to the single orthologous locus with a threshold of at least 95% sequence identity;

(2) During the reciprocal LiftOver test, the NHP sequence identified in the direct LiftOver test has been remapped with at least 95% sequence identity threshold to the exactly same human orthologous sequence which was queried during the direct LiftOver test.

A set of experimentally defined 3583 regulatory loci of the mouse ESC which are documented to harbor TFBS for multiple TFs (Chen et al., 2008) was utilized as a reference to estimate changes of TF-binding patterns and the TFBS densities during mammalian evolution. To identify TFBS within genomic regulatory loci known to contribute to regulation of human and chimpanzee brain development, a total of 17,935 genomic regulatory regions reported in the organoid single-cell genomic atlas of human and chimpanzee brain development (Kanton et al, 2019) were analyzed. A catalogue of TFs having putative TFBS within 8099 human-specific and 9836 chimpanzee-specific brain development regulatory regions (BDRRs) was compiled. BDRRs were identified employing DNA sequence unbiased open chromatin accessibility screening method termed the Assay for Transposase-Accessible Chromatin with high-throughput sequencing (ATAC-seq; Buenrostro et al, 2013; 2015). A set of 751 TE loci of the SVA families coopted as functional cis-regulatory elements in human induced pluripotent stem cells (Barnada et al., 2022) and a set of 3,265 TE loci engaged in TE-governed hippocampal neurogenesis regulatory pathways of human and chimpanzee (Patoori et al., 2022) were analyzed. TE-governed hippocampal neurogenesis regulatory pathways were discovered by Patoori et al. (2022) utilizing a model of differentiation of human and chimpanzee induced pluripotent stem cells into TBR2 (or EOMES)-positive hippocampal intermediate progenitor cells.

Identification of transcription factor binding sites (TFBS) within candidate genomic regulatory loci was performed employing the Jaspar algorithm (Jaspar Transcription Factors) accessible through the UCSC Genome Bowser Table Browser functions ( https://genome.ucsc.edu/cgi-bin/hgTables ) facilitating downloading, filtering, analyzing, and retrieving data from the Genome Browser. TFBS identification and data retrieval were carried out using the default thresholds settings of imputing up to 1,000 loci per screen to achieve the full coverage of the specified genomic regions of interest based on coordinated of hg38 and hg19 human reference genome databases. All identified TFBS were retrieved and all individual TFs having TFBS were catalogued. For each set of TFBS, quantitative features were documented by calculating the numbers of events recorded for each distinct TFBS and reporting the computed values as numbers of TFBS per regulatory locus (defined as TFBS frequency) and the estimated TFBS density calculated as TFBS frequency normalized to 1 Kb of the locus length (defined as TFBS density).

The significance of the differences in the expected and observed numbers of events was calculated using two-tailed Fisher's exact test. Multiple proximity placement enrichment tests were performed for individual families and sub-sets of LTRs, BDRRs, and human-specific regulatory regions (HSRS) taking into account the size in bp of corresponding genomic regions, size distributions in human cells of topologically associating domains, distances to putative regulatory targets, bona fide regulatory targets identified in targeted genetic interference and/or epigenetic silencing experiments. Additional details of methodological and analytical approaches are provided in the text, Supplementary Materials and previously reported contributions [Barakat et al. 2018; Fuentes et al. 2018; Glinsky 2015; 2016a, b; 2018; 2019; 2020a, b, c, 2021; 2022;

Guffanti et al. 2018; Glinsky and Barakat, 2019; McLean et al. 2010; 2011; Pontis et al. 2019; Wang et al. 2014].

*Gene set enrichment and genome-wide proximity placement analyses*

Gene set enrichment analyses were carried-out using the Enrichr bioinformatics platform, which enables the interrogation of nearly 200,000 gene sets from more than 100 gene set libraries. The Enrichr API (January 2018 through January 2023 releases) [Chen et al. 2013; Kuleshov et al. 2016; Xie et al. 2021] was used to test genes linked to regulatory LTR elements, HSRS, or other regulatory loci of interest for significant enrichment in numerous functional categories. When technically feasible, larger sets of genes comprising several thousand entries were analyzed. Regulatory connectivity maps between HSRS, regulatory LTRs and coding genes and additional functional enrichment analyses were performed with the Genomic Regions Enrichment of Annotations Tool (GREAT) algorithm [McLean et al. 2010; 2011] at default settings. The reproducibility of the results was validated by implementing two releases of the GREAT algorithm: GREAT version 3.0.0 (02/15/2015 to 08/18/2019) and GREAT version 4.0.4 (08/19/2019) applying default settings at differing maximum extension thresholds as previously reported (Glinsky 2020a, b, c; 2021; 2022; 2024). The GREAT algorithm allows investigators to identify and annotate the genome-wide connectivity networks of user-defined distal regulatory loci and their putative target genes. Concurrently, the GREAT algorithm performs functional Gene Ontology (GO) annotations and analyses of statistical enrichment of GO annotations of identified genomic regulatory elements (GREs) and target genes, thus enabling the inference of potential biological impacts of interrogated genomic regulatory networks. The Genomic Regions Enrichment of Annotations Tool (GREAT) algorithm was employed to identify putative down-stream target genes of human embryo regulatory LTRs. Concurrently with the identification of putative regulatory target genes of GREs, the GREAT algorithm performs stringent statistical enrichment analyses of functional annotations of identified down-stream target genes, thus enabling the inference of potential significance of phenotypic impacts of interrogated GRNs. Importantly, the assignment of phenotypic traits as putative statistically valid components of GRN actions entails the assessments of statistical significance of the enrichment of both GREs and down-stream target genes by applying independent statistical tests.

The validity of statistical definitions of genomic regulatory networks (GRNs) and genomic regulatory modules (GRMs) based on the binominal (regulatory elements) and hypergeometrc (target genes) FDR Q values was evaluated using a directed acyclic graph (DAG) test based on the enriched terms from a single ontology-specific table generated by the GREAT algorithm (Glinsky, 2024). DAG test draws patterns and directions of connections between significantly enriched GO modules based on the experimentally-documented temporal logic of developmental processes and structural/functional relationships between gene ontology enrichment analysis-defined statistically significant terms. A specific DAG test utilizes only a sub-set of statistically significant GRMs from a single gene ontology-specific table generated by the GREAT algorithm by extracting GRMs manifesting connectivity patterns defined by experimentally documented developmental and/or structure/function/activity relationships. These GRMs are deemed valid observations and visualized as a consensus hierarchy network of the ontology-specific DAGs (Glinsky, 2024). Based on these considerations, the DAG algorithm draws the developmental and structure/function/activity relationships-guided hierarchy of connectivity between statistically significant gene ontology enriched GRMs.

Genome-wide Proximity Placement Analysis (GPPA) of down-stream target genes and distinct genomic features co-localizing with regulatory LTRs, HSRS, BDDRs and other regulatory loci was carried out as described previously and originally implemented for human-specific transcription factor binding sites [Glinsky et al. 2018; Glinsky, 2015, 2016a, 2016b, 2017, 2018, 2019, 2020a, 2020b, 2020c, 2021; 2022; 2024; Guffanti et al. 2018].

*Differential GSEA to infer the relative contributions of distinct subsets of regulatory LTR elements and down-stream target genes on phenotypes of interest.*

When technically and analytically feasible, different sets of regulatory LTRs and candidate down-stream target genes defined at several significance levels of statistical metrics and comprising from dozens to several thousand individual genetic loci were analyzed using differential GSEA. This approach was utilized to gain insights into biological effects of regulatory LTRs and down-stream target genes and infer potential mechanisms of phenotype affecting activities. Previously, this approach was successfully implemented for identification and characterization of human-specific regulatory networks governed by human-specific transcription factor-binding sites [Glinsky et al. 2018; Glinsky, 2015, 2016a, 2016b, 2017, 2018, 2019, 2020a,

2020b, 2020c, 2021; 2022; 2024; Guffanti et al. 2018] and functional enhancer elements [Barakat et al. 2018; Glinsky et al. 2018; Glinsky and Barakat 2019; Glinsky 2015, 2016a, 2016b, 2017, 2018, 2019, 2020a, 2020b, 2020c, 2021; 2022; 2024]. Differential GSEA approach has been utilized for characterization of phenotypic impacts of 13,824 genes associated with 59,732 human-specific regulatory sequences [Glinsky, 2020a], 8,405 genes associated with 35,074 human-specific neuroregulatory single-nucleotide changes [Glinsky, 2020b], 8,384 genes regulated by stem cell-associated retroviral sequences (SCARS) [Glinsky 2021], as well as human genes and medicinal molecules affecting the susceptibility to SARS-CoV-2 coronavirus [Glinsky, 2020c].

Initial GSEA entail interrogations of each specific set of candidate down-stream target genes using ~70 distinct genomic databases, including comprehensive pathway enrichment Gene Ontology (GO) analyses. Upon completion, these analyses were followed by in-depth interrogations of the identified significantly-enriched gene sets employing selected genomic databases deemed most statistically informative at the initial GSEA. In all reported tables and plots (unless stated otherwise), in addition to the nominal p values and adjusted p values, the Enrichr software calculate the "combined score", which is a product of the significance estimate and the magnitude of enrichment (combined score $c = \log(p) * z$, where p is the Fisher's exact test p-value and z is the z-score deviation from the expected rank).

*Statistical Analyses of the Publicly Available Datasets*

All statistical analyses of the publicly available genomic datasets, including error rate estimates, background and technical noise measurements and filtering, feature peak calling, feature selection, assignments of genomic coordinates to the corresponding builds of the reference human genome, and data visualization, were performed exactly as reported in the original publications and associated references linked to the corresponding data visualization tracks (http://genome.ucsc.edu/ ). Additional elements or modifications of statistical analyses are described in the corresponding sections of the Results. Statistical significance of the Pearson correlation coefficients was determined using GraphPad Prism version 6.00 software. Both nominal and Bonferroni adjusted p values were estimated and considered as reported in corresponding sections of the Results. The significance of the differences in the numbers of events between the groups was calculated using two-sided Fisher's exact and Chi-square test, and the significance of the overlap between the events was determined using the hypergeometric distribution test [Tavazoie et al. 1999].

**Supplementary Information is available online.**

Supplementary information includes Supplementary Tables S1-S6; Supplemenmtary Figure S1; and Supplementary Summaries S1-S3.

**Acknowledgements.** This work was made possible by the open public access policies of major grant funding agencies and international genomic databases and the willingness of many investigators worldwide to share their primary research. Author would like to thank you Victoria Glinskii for invaluable expert assistance with graphical presentation of the results of this study.

**Author Contributions**

This is a single author contribution. All elements of this work, including the conception of ideas, formulation, and development of concepts, execution of experiments, analysis of data, and writing of the paper, were performed by the author.

**Funding**

In part, this work was supported by OncoScar, LLC.

**Declarations**

**Conflict of interest statement**

No conflicts of interest to declare.

**Data availability statement**

All data supporting the reported observations and required to reproduce the findings are provided in the main body of the paper and Supplementary materials.

**Ethics approval and consent to participate**

Not applicable

**Consent for publication**

Not applicable

**Figure legends**

**Figure 1.** DNA sequences of four distinct families of human embryo regulatory LTRs (LTR5_Hs; MLT2A1; MLT2A2; LTR7) harbor nearly identical arrays of ~770 transcription factor binding sites (TFBS) for 716 proteins, a vast majority of which (770 of 716; 98%) comprises protein constituents of these networks appear to share common Gene Ontology (GO) features of sequence-specific double-stranded DNA binding (GO: 1990837).

A. Graphical representations of identity profiles of TFBS harbored by four distinct families of human embryo regulatory LTRs (LTR5_Hs; MLT2A1; MLT2A2; LTR7).

B. Correlation patterns of TFBS density profiles of human MLT2A2 loci and multiTF-binding loci of mouse ESC.

C. Correlation patterns of TFBS density profiles of human MLT2A1 loci and multiTF-binding loci of mouse ESC.

D. Correlation patterns of TFBS density profiles of human LTR7 loci and multiTF-binding loci of mouse ESC.

E. Correlation patterns of TFBS density profiles of human LTR5_Hs loci and multiTF-binding loci of mouse ESC.

TFBS densities were computed for 100 top-scoring TFs having TFBS within 606 LTR5_Hs loci; as well as MLT2A1; MLT2A2; LTR7 regulatory elements and aligned to TFBS density profiles estimated for 3583 mouse ESC multi-TFs-binding regulatory loci (see text for details and Supplementary Tables S1 and S2 for additional information).

**Figure 2.** Graphical summary of the analyses of TFBS density gains and losses acquired by regulatory loci of four families of human embryo regulatory LTRs.

A. Transcription factors manifesting largest TFBS density gains compared with mouse ESC multi-TFs-binding loci.

B. Transcription factors of human embryo LTRs manifesting more than 50% gain/loss of TFBS density compared with mouse ESC multi-TFs-binding loci and brief summary of GSEA 30 TFs of human embryo LTRs manifesting more than 50% gain of TFBS density.

C. Correlation patterns of TFBS density profiles of human LTR5_Hs and LTR7 loci for 100 top-scoring TFs.

D. Correlation patterns of TFBS density profiles of human LTR5_Hs and LTR7 loci for 563 top-scoring TFs.

E-F. LTR5_Hs transcription factors manifesting largest TFBS density gains compared with LTR7, MLT2A1, and MLT2A2 loci at 100% gain threshold (E) and 50% gain threshold (F).

G-I. A brief summary of GSEA of 44 LTR5_Hs TFs manifesting more than 50% gain of TFBS density in comparisons among different families of human embryo regulatory LTRs.

TFBS densities were computed for 100 top-scoring TFs having TFBS within 606 LTR5_Hs loci; as well as MLT2A1; MLT2A2; LTR7 regulatory elements and aligned to TFBS density profiles estimated for 3583 mouse ESC multi-TFs-binding regulatory loci to estimate gains and/or losses relative to corresponding TFBS density values of mouse ESC multi-TFs-binding regulatory loci (see text for details and Supplementary Tables S1-S4 for additional information).

**Figure 3.** Genome-wide similarity and divergence matrix of TFBS density changes acquired during mammalian evolution within 17,935 brain development regulatory regions (BDRRs) on individual chromosomes of human and chimpanzee.

A. Similarity matrix of TFBS gains/losses within brain development regulatory regions acquired on individual chromosomes by human and chimpanzee during 80 MYR of mammalian evolution.

B. Common patterns of TFBS density changes within BDRRs acquired on chromosome 19 by Chimpanzee and Modern Humans.

C. Common patterns of TFBS density changes within BDRRs acquired on chromosome 22 by Chimpanzee and Modern Humans.

D. Distinct patterns of TFBS density changes within BDRRs acquired on chromosome 14 by Chimpanzee and Modern Humans.

E. Distinct patterns of TFBS density changes within BDRRs acquired on chromosome 21 by Chimpanzee and Modern Humans.

See text and Table 2 for details and additional information.

**Figure 4.** Genome-wide chromosome-level visualization of distinct patterns of TFBS density gains and losses within 17,935 brain development regulatory regions of Modern Humans and Chimpanzee acquired during mammalian evolution.

A. Divergence patterns of TFBS density gains in Human and Chimpanzee regulatory regions of brain development during mammalian evolution.

B. Divergence patterns of TFBS density loss in Human and Chimpanzee regulatory regions of brain development during mammalian evolution.

C. Divergence patterns of TFBS density gains in Human versus Chimpanzee regulatory regions of brain development.

D. Divergent patterns of TFBS density changes of at least 25% acquired by Chimpanzee and Modern Humans during primate evolution after segregation from last common ancestor.

TFBS densities were computed for BDRRs located on individual chromosomes of human and chimpanzee genomes and for each TF gain or loss of TFBS density were estimated by comparisons to corresponding values of TFBS densities of mouse ESC multi-TFs-binding regulatory loci and reported as numbers of gains (A) and losses (B) of at least 25%. Panels (C) and (D) show results of the direct TFBS density comparisons within BDRRs of human versus chimpanzee.

Results of the analyses of gains and losses of TFBS densities are reported for 100 top-scoring TFs as defined by the analyses of TFBS within LTR5_Hs loci (Supplementary Tables S1-S4).

**Figure 5.** Patterns of divergence and concordance within human and chimpanzee genomes ascertained from genome-wide chromosome-level correlation matrices of TFBS density changes acquired during mammalian evolution within Modern Humans (A; B) and Chimpanzee (C; D) brain development regulatory regions defined by the ATAC-seq analysis.

A. Visualization of human genome-wide divergence scores ascertained from genome-wide chromosome-level correlation matrix reported in the Table 3.

B. Visualization of human genome-wide variation coefficients ascertained from genome-wide chromosome-level correlation matrix reported in the Table 3.

C. Visualization of chimpanzee genome-wide divergence scores ascertained from genome-wide chromosome-level correlation matrix reported in the Table 4.

D. Visualization of chimpanzee genome-wide variation coefficients ascertained from genome-wide chromosome-level correlation matrix reported in the Table 4.

Genome-wide chromosome-level correlation matrices of TFBS density changes acquired during mammalian evolution within Modern Humans (Table 3) and Chimpanzee (Table 4) brain development regulatory regions defined by the ATAC-seq analysis were created and utilized for computation of divergence scores and variation coefficients for individual chromosomes. See text for details.

**Figure 6.** Distinct and common association patterns of TFBS density changes within brain development regulatory regions of human and chimpanzee revealed by genome-wide chromosome level alignment analyses with TFBS density profiles of different families of human embryo regulatory LTRs.

A. Chromosome-level correlation patterns of TFBS density changes acquired during mammalian evolution within human and chimpanzee brain development regulatory regions aligned to MLT2A1 loci.

B. Chromosome-level correlation patterns of TFBS density changes acquired during mammalian evolution within human and chimpanzee brain development regulatory regions aligned to LTR5_Hs loci.

C. Chromosome-level correlation patterns of TFBS density changes acquired during mammalian evolution within human and chimpanzee brain development regulatory regions aligned to MLT2A2 loci.

D. Chromosome-level correlation patterns of TFBS density changes acquired during mammalian evolution within human and chimpanzee brain development regulatory regions aligned to LTR7 loci.

E-F. Chromosome-level discordant correlation patterns of TFBS density changes acquired during mammalian evolution within human (E) and chimpanzee (F) brain development regulatory regions aligned to LTR5_Hs and MLT2A1 loci.

G-H. Chromosome-level negative correlation patterns of TFBS density changes acquired during mammalian evolution within human (G) and chimpanzee (H) brain development regulatory regions aligned to LTR5_Hs and MLT2A1 loci.

I-K. Interspecies chromosome-level correlation profiles of TFBS density changes acquired during mammalian evolution within human and chimpanzee brain development regulatory regions aligned to MLT2A1 loci (I); or to LTR5_Hs loci (K); and combining alignment patterns to both MLT2A1 and LTR5_Hs loci (L).

L. Common patterns of genome-wide TFBS density changes acquired by Chimpanzee and Modern Humans during primate evolution identified by comparisons to TFBS density values within mouse ESC multi-TFs-binding regulatory loci.

M. Divergent patterns of TFBS density changes of at least 25% acquired by Chimpanzee and Modern Humans during primate evolution identified by direct comparisons of TFBS density gains and losses within BDRRs of human versus chimpanzee. Highlighted are changes of at least 45%.

**Figure 7.** DNA sequences of LTR5_Hs subfamily of human embryo regulatory LTRs and five distinct families of transposable elements (TE) coopted in genomic regulatory networks of differentiation of human and chimpanzee induced pluripotent stem cells into TBR2 (or EOMES)-positive hippocampal intermediate progenitor cells harbor nearly identical arrays of ~770 transcription factor binding sites (TFBS) for 716 proteins, a vast majority of which (770 of 716; 98%) comprises protein constituents of these networks appear to share common Gene Ontology (GO) features of sequence-specific double-stranded DNA binding (GO: 1990837).

A. Graphical representations of identity profiles of TFBS harbored by LTR5_Hs subfamily of human embryo regulatory LTRs and five distinct families of TE implicated in genomic regulatory networks of differentiation of human and chimpanzee induced pluripotent stem cells into TBR2 (or EOMES)-positive hippocampal intermediate progenitor cells.

B. Graphical representations of identity profiles of TFBS harbored by LTR5_Hs subfamily of human embryo regulatory LTRs and six distinct subfamilies of LINE1 retrotransposons implicated in genomic regulatory networks of differentiation of human and chimpanzee induced pluripotent stem cells into TBR2 (or EOMES)-positive hippocampal intermediate progenitor cells.

C. Diminished representation of divergent LTR loci within differentially-accessed (DA) ATAC regulatory regions of differentiation of human induced pluripotent stem cells into TBR2 (or EOMES)-positive hippocampal intermediate progenitor cells.

D. Diminished representation of divergent SINE/Alu loci within differentially-accessed (DA) ATAC regulatory

regions of differentiation of human induced pluripotent stem cells into TBR2 (or EOMES)-positive hippocampal intermediate progenitor cells.

E. DA ATAC regulatory regions of differentiation of human induced pluripotent stem cells into TBR2 (or EOMES)-positive hippocampal intermediate progenitor cells intersecting high numbers of SINE/Alu loci house more transcription factors with TFBS.

F. Diminished representation of divergent LINE loci within differentially-accessed (DA) ATAC regulatory regions of differentiation of human induced pluripotent stem cells into TBR2 (or EOMES)-positive hippocampal intermediate progenitor cells.

TFs identities were determined and TFBS densities were computed for 100 top-scoring TFs having TFBS within 606 LTR5_Hs loci as well as TE-encoded regulatory elements coopted in genomic regulatory networks of differentiation of human and chimpanzee induced pluripotent stem cells into TBR2 (or EOMES)-positive hippocampal intermediate progenitor cells.

**Figure 8.** Distinct and common association patterns of TFBS density changes within brain development regulatory regions of human and chimpanzee revealed by genome-wide chromosome level alignment analyses with TFBS density profiles of distinct families of TE coopted in genomic regulatory networks of differentiation of human and chimpanzee induced pluripotent stem cells into TBR2 (or EOMES)-positive hippocampal intermediate progenitor cells.

A. Chromosome-level correlation patterns of TFBS density changes acquired during mammalian evolution within human and chimpanzee brain development regulatory regions aligned to SVA loci.

B. Chromosome-level discordant correlation patterns of TFBS density changes acquired during mammalian evolution within human brain development regulatory regions aligned to SVA and MLT2A1 loci.

C. Chromosome-level discordant correlation patterns of TFBS density changes acquired during mammalian evolution within chimpanzee brain development regulatory regions aligned to SVA and MLT2A1 loci.

D. Chromosome-level correlation patterns of TFBS density changes acquired during mammalian evolution within human and chimpanzee brain development regulatory regions aligned to L1PA6 loci.

E-F. Chromosome-level discordant correlation patterns of TFBS density changes acquired during mammalian evolution within human (E) and chimpanzee (F) brain development regulatory regions aligned to L1PA6 and SVA loci.

G-H. Chromosome-level negative correlation patterns of TFBS density changes acquired during mammalian evolution within human (G) and chimpanzee (H) brain development regulatory regions aligned to LTR5_Hs and L1PA6 loci.

I-J. Chromosome-level positive correlation patterns of TFBS density changes acquired during mammalian evolution within human (I) and chimpanzee (K) brain development regulatory regions aligned to MLTA1 and L1PA6 loci.

K-L. Chromosome-level negative correlation patterns of TFBS density changes acquired during mammalian evolution within human (L) and chimpanzee (M) brain development regulatory regions aligned to SVA and L1PA6 loci.

M-O. Interspecies chromosome-level correlation profiles of TFBS density changes acquired during mammalian evolution within human and chimpanzee brain development regulatory regions aligned to SVA loci (N); or to L1PA6 loci (O); and combining alignment patterns to both SVA and L1PA6 loci (L).

**Table 1.** Summary of the GSEA of 716 genes having 771 distinct TFBS within LTR5_Hs loci defined as the Sequence-Specific Double-Stranded DNA Binding pathway.

| Database | Number of records* | Representative top-scoring record | Number of genes | Adjusted p value |
|---|---|---|---|---|
| **Cells and tissues** | | | | |
| Allen Brain Atlas down | 145 | Intermediate part of isB | 34 | 3.46E-06 |
| Allen Brain Atlas up | 654 | Primary visual area | 33 | 5.41E-06 |
| ARCHS4 Tissues | 40 | Amniotic fluid | 151 | 9.42E-12 |
| Jensen Tissues | 130 | Finger | 99 | 1.20E-48 |
| **Single cells** | | | | |
| CellMarker 2024 | 276 | Neural Stem Cell Undefined Human | 18 | 1.31E-13 |
| CellMarker Augmented 2021 | 70 | Neural Stem cell: Undefined | 19 | 5.49E-11 |
| Descartes Cell Types and Tissue 2021 | 10 | Inhibitory neurons in Cerebrum | 14 | 2.41E-08 |
| PanglaoDB Augmented 2021 | 36 | Motor Neurons | 39 | 3.17E-26 |
| **Diseases** | | | | |
| DisGeNET | 1479 | Carcinogenesis | 314 | 3.58E-44 |
| Orphanet Augmented 2021 | 156 | Brachydactyly type D ORPHA:93385** | 56 | 1.55E-51 |
| Rare Diseases AutoRIF Gene Lists | 1184 | Branchial arch defects | 63 | 1.79E-49 |
| Rare Diseases GeneRIF Gene Lists | 730 | Embryonal carcinoma | 49 | 3.01E-18 |
| GTEx Aging Signatures 2021 | 15 | GTEx Blood Vessels 20-29 vs 60-69 Down | 26 | 1.60E-04 |
| **Protein-protein interactions (PPI)** | | | | |
| PPI Hub Proteins | 73 | CREBBP | 139 | 1.14E-110 |
| Transcription Factor PPIs | 205 | EP300 | 161 | 7.20E-113 |
| Virus-Host PPI P-HIPSTer 2020 | 69 | Human papillomavirus type 6 E5B (gene: E5B) | 69 | 4.72E-29 |
| **Gene ontology** | | | | |
| GO Biological Process 2023 | 591 | Regulation Of Transcription By RNA Polymerase II (GO:0006357) | 628 | 0.00E+00 |
| GO Cellular Component 2023 | 4 | Intracellular Membrane-Bounded Organelle (GO:0043231) | 525 | 2.91E-160 |
| GO Molecular Function 2023 | 53 | RNA Polymerase II Transcription Regulatory Region Sequence-Specific DNA Binding (GO:0000977) | 609 | 0.00E+00 |
| MGI Mammalian Phenotype Level 4 2021 | 1613 | Neonatal lethality, complete penetrance MP:0011087 | 99 | 3.75E-44 |
| Human Phenotype Ontology | 199 | Autosomal dominant inheritance (HP:0000006) | 126 | 8.02E-28 |
| Jensen Compartments | 136 | Nucleoplasm | 292 | 4.01E-65 |
| **Regulatory pathways** | | | | |
| ChEA 2022 | 489 | BMI1 23680149 ChIP-Seq NPCs Mouse | 228 | 1.87E-137 |
| ENCODE and ChEA Consensus TFs from ChIP-X | 35 | SUZ12 CHEA | 284 | 1.16E-120 |
| Elsevier Pathway Collection | 561 | TERT Activation in Cancer | 21 | 1.38E-18 |
| KEGG 2021 Human | 76 | Transcriptional misregulation in cancer | 55 | 3.54E-32 |
| Reactome 2022 | 180 | RNA Polymerase II Transcription R-HSA-73857 | 180 | 2.14E-56 |
| The Kinase Library 2023 | 59 | JNK1 | 63 | 1.00E-07 |

Legend: *, Number of significantly-enriched records defined at the adjusted p-value threshold < 0.05; **, Non rare in Europe; Comprehensive reports of the Gene Set Enrichment Analyses (GSEA) are presented in the Supplementary Table S1, including statistical metrics (see Methods) and complete lists of genes linked to each significantly enriched records.

**Table 2.** Genome-wide comparative analyses of the Sequence-Specific Double-Stranded DNA Binding pathway within ATAC-seq*-defined brain development regulatory regions of Modern Humans and Chimpanzee.

| Chromosome | Species | TFBS number | ATAC span, bp | Number of ATAC DA loci | Per ATAC DA locus | Mean locus size | TFBS density per 1 Kb | TFs number | Species | TFBS number | ATAC span, bp | Number of ATAC DA loci | Per ATAC DA locus | Mean locus size | TFBS density per 1 Kb | Human gain, % TFBS density | TFs number |
|---|---|---|---|---|---|---|---|---|---|---|---|---|---|---|---|---|---|
| chr1 | Chimp | 1766356 | 446722 | 830 | 2128 | 538 | 3956 | 776 | Human | 1526024 | 355383 | 670 | 2278 | 530 | 4297 | 8.6 | 776 |
| chr2 | Chimp | 1695427 | 407699 | 758 | 2237 | 538 | 4157 | 776 | Human | 1836793 | 415038 | 783 | 2346 | 530 | 4426 | 6.5 | 776 |
| chr3 | Chimp | 1604879 | 380914 | 702 | 2286 | 543 | 4210 | 776 | Human | 1234478 | 273715 | 517 | 2388 | 529 | 4514 | 7.2 | 776 |
| chr4 | Chimp | 1625042 | 365446 | 689 | 2359 | 530 | 4450 | 776 | Human | 1125139 | 245659 | 467 | 2409 | 526 | 4580 | 2.9 | 776 |
| chr5 | Chimp | 1321485 | 308669 | 572 | 2310 | 540 | 4278 | 776 | Human | 1357153 | 300706 | 560 | 2423 | 537 | 4513 | 5.5 | 776 |
| chr6 | Chimp | 1420573 | 335204 | 619 | 2295 | 542 | 4234 | 775 | Human | 1250703 | 279891 | 529 | 2364 | 529 | 4469 | 5.6 | 776 |
| chr7 | Chimp | 1347657 | 323571 | 589 | 2288 | 549 | 4168 | 776 | Human | 968451 | 217196 | 402 | 2409 | 540 | 4461 | 7.0 | 776 |
| chr8 | Chimp | 1093119 | 257416 | 475 | 2301 | 542 | 4246 | 776 | Human | 912095 | 205404 | 387 | 2357 | 531 | 4438 | 4.5 | 776 |
| chr9 | Chimp | 712088 | 175427 | 319 | 2232 | 550 | 4059 | 776 | Human | 842829 | 192780 | 367 | 2297 | 525 | 4374 | 7.8 | 775 |
| chr10 | Chimp | 917165 | 220274 | 407 | 2253 | 541 | 4165 | 774 | Human | 868733 | 202055 | 386 | 2251 | 523 | 4303 | 3.3 | 776 |
| chr11 | Chimp | 1182974 | 293637 | 538 | 2199 | 546 | 4027 | 776 | Human | 871740 | 201055 | 387 | 2253 | 520 | 4332 | 7.6 | 775 |
| chr12 | Chimp | 1024868 | 249740 | 459 | 2233 | 544 | 4104 | 776 | Human | 982916 | 221742 | 412 | 2386 | 538 | 4434 | 8.0 | 776 |
| chr13 | Chimp | 682310 | 155878 | 291 | 2345 | 536 | 4374 | 776 | Human | 689718 | 150644 | 282 | 2446 | 534 | 4580 | 4.7 | 774 |
| chr14 | Chimp | 659570 | 162339 | 294 | 2243 | 552 | 4064 | 776 | Human | 670324 | 152454 | 290 | 2311 | 526 | 4394 | 8.1 | 776 |
| chr15 | Chimp | 550681 | 139607 | 254 | 2168 | 550 | 3942 | 776 | Human | 470808 | 112541 | 211 | 2231 | 533 | 4186 | 6.2 | 775 |
| chr16 | Chimp | 556923 | 146086 | 275 | 2025 | 531 | 3814 | 774 | Human | 469037 | 113960 | 223 | 2103 | 511 | 4116 | 7.9 | 774 |
| chr17 | Chimp | 691939 | 189546 | 350 | 1977 | 542 | 3648 | 775 | Human | 476570 | 120417 | 233 | 2045 | 517 | 3956 | 8.5 | 774 |
| chr18 | Chimp | 538080 | 125611 | 232 | 2319 | 541 | 4287 | 774 | Human | 607767 | 137539 | 257 | 2365 | 535 | 4420 | 3.1 | 775 |
| chr19 | Chimp | 472720 | 139234 | 250 | 1891 | 557 | 3395 | 783 | Human | 288627 | 79278 | 151 | 1911 | 525 | 3641 | 7.2 | 774 |
| chr20 | Chimp | 473627 | 121465 | 222 | 2133 | 547 | 3900 | 775 | Human | 369727 | 87944 | 168 | 2201 | 523 | 4208 | 7.9 | 775 |
| chr21 | Chimp | 254160 | 62717 | 119 | 2136 | 527 | 4053 | 773 | Human | 197724 | 46450 | 88 | 2247 | 528 | 4255 | 5.0 | 771 |
| chr22 | Chimp | 287741 | 79889 | 148 | 1944 | 540 | 3600 | 773 | Human | 217597 | 57388 | 108 | 2015 | 531 | 3794 | 5.4 | 773 |
| chrX | Chimp | 967044 | 240330 | 444 | 2178 | 541 | 4026 | 776 | Human | 500381 | 114863 | 221 | 2264 | 520 | 4354 | 8.2 | 772 |
| Genome | Chimp | 21846428 | 5327421 | 9836 | 2221 | 542 | 4098 | 776 | Human | 18735334 | 4284102 | 8099 | 2313 | 529 | 4373 | 6.7 | 775 |

| P-value** |
|---|
| 0.001501 |
| 5.61E-05 |
| 0.000169 |
| 8.89E-10 |
| 2.67E-07 |
| 9.35E-15 |
| 0.11465 |

Legends: *, ATAC-seq designates the Assay for Transposase-Accessible Chromatin with high-throughput sequencing; **, 2-tail t-test (paired);

**Table 3.** Genome-wide chromosome-level correlation matrix of TFBS density changes acquired during mammalian evolution within human brain development regulatory regions defined by the ATAC-seq analysis.

| Chromosome | chr1 | chr2 | chr3 | chr4 | chr5 | chr6 | chr7 | chr8 | chr9 | chr10 | chr11 | chr12 |
|---|---|---|---|---|---|---|---|---|---|---|---|---|
| chr1 | **1.000** | 0.751 | 0.643 | 0.680 | 0.671 | 0.697 | 0.785 | 0.699 | 0.887 | 0.904 | 0.921 | 0.844 |
| chr2 | 0.751 | **1.000** | 0.965 | 0.974 | 0.968 | 0.973 | 0.972 | 0.967 | 0.937 | 0.935 | 0.920 | 0.965 |
| chr3 | 0.643 | 0.965 | **1.000** | 0.993 | 0.992 | 0.991 | 0.972 | 0.986 | 0.897 | 0.890 | 0.866 | 0.947 |
| chr4 | 0.680 | 0.974 | 0.993 | **1.000** | 0.996 | 0.995 | 0.982 | 0.991 | 0.916 | 0.915 | 0.887 | 0.956 |
| chr5 | 0.671 | 0.968 | 0.992 | 0.996 | **1.000** | 0.997 | 0.981 | 0.993 | 0.917 | 0.913 | 0.886 | 0.950 |
| chr6 | 0.697 | 0.973 | 0.991 | 0.995 | 0.997 | **1.000** | 0.986 | 0.993 | 0.929 | 0.926 | 0.902 | 0.961 |
| chr7 | 0.785 | 0.972 | 0.972 | 0.982 | 0.981 | 0.986 | **1.000** | 0.981 | 0.962 | 0.961 | 0.945 | 0.984 |
| chr8 | 0.699 | 0.967 | 0.986 | 0.991 | 0.993 | 0.993 | 0.981 | **1.000** | 0.925 | 0.923 | 0.902 | 0.955 |
| chr9 | 0.887 | 0.937 | 0.897 | 0.916 | 0.917 | 0.929 | 0.962 | 0.925 | **1.000** | 0.987 | 0.989 | 0.973 |
| chr10 | 0.904 | 0.935 | 0.890 | 0.915 | 0.913 | 0.926 | 0.961 | 0.923 | 0.987 | **1.000** | 0.986 | 0.974 |
| chr11 | 0.921 | 0.920 | 0.866 | 0.887 | 0.886 | 0.902 | 0.945 | 0.902 | 0.989 | 0.986 | **1.000** | 0.965 |
| chr12 | 0.844 | 0.965 | 0.947 | 0.956 | 0.950 | 0.961 | 0.984 | 0.955 | 0.973 | 0.974 | 0.965 | **0.473** |
| chr13 | 0.651 | 0.961 | 0.992 | 0.993 | 0.994 | 0.994 | 0.977 | 0.986 | 0.905 | 0.897 | 0.873 | 0.945 |
| chr14 | 0.764 | 0.977 | 0.972 | 0.981 | 0.983 | 0.986 | 0.987 | 0.983 | 0.959 | 0.954 | 0.940 | 0.973 |
| chr15 | 0.971 | 0.751 | 0.648 | 0.688 | 0.686 | 0.708 | 0.785 | 0.706 | 0.897 | 0.909 | 0.920 | 0.833 |
| chr16 | 0.803 | 0.248 | 0.084 | 0.132 | 0.121 | 0.159 | 0.280 | 0.164 | 0.470 | 0.504 | 0.538 | 0.380 |
| chr17 | 0.530 | -0.111 | -0.288 | -0.247 | -0.263 | -0.225 | -0.098 | -0.221 | 0.104 | 0.138 | 0.183 | 0.017 |
| chr18 | 0.723 | 0.991 | 0.957 | 0.968 | 0.962 | 0.966 | 0.958 | 0.958 | 0.923 | 0.924 | 0.900 | 0.949 |
| chr19 | 0.286 | -0.386 | -0.536 | -0.500 | -0.510 | -0.478 | -0.358 | -0.473 | -0.158 | -0.128 | -0.081 | -0.249 |
| chr20 | 0.948 | 0.618 | 0.496 | 0.534 | 0.531 | 0.557 | 0.648 | 0.552 | 0.793 | 0.807 | 0.825 | 0.727 |
| chr21 | 0.787 | 0.960 | 0.943 | 0.958 | 0.952 | 0.958 | 0.964 | 0.963 | 0.939 | 0.945 | 0.925 | 0.961 |
| chr22 | 0.383 | -0.283 | -0.447 | -0.399 | -0.410 | -0.375 | -0.253 | -0.377 | -0.044 | -0.013 | 0.028 | -0.156 |
| chrX | 0.845 | 0.945 | 0.930 | 0.940 | 0.941 | 0.947 | 0.972 | 0.954 | 0.967 | 0.971 | 0.963 | 0.978 |

| | chr13 | chr14 | chr15 | chr16 | chr17 | chr18 | chr19 | chr20 | chr21 | chr22 | chrX |
|---|---|---|---|---|---|---|---|---|---|---|---|
| chr1 | 0.651 | 0.764 | 0.971 | 0.803 | 0.530 | 0.723 | 0.286 | 0.948 | 0.787 | 0.383 | 0.845 |
| chr2 | 0.961 | 0.977 | 0.751 | 0.248 | -0.111 | 0.991 | -0.386 | 0.618 | 0.960 | -0.283 | 0.945 |
| chr3 | 0.992 | 0.972 | 0.648 | 0.084 | -0.288 | 0.957 | -0.536 | 0.496 | 0.943 | -0.447 | 0.930 |
| chr4 | 0.993 | 0.981 | 0.688 | 0.132 | -0.247 | 0.968 | -0.500 | 0.534 | 0.958 | -0.399 | 0.940 |
| chr5 | 0.994 | 0.983 | 0.686 | 0.121 | -0.263 | 0.962 | -0.510 | 0.531 | 0.952 | -0.410 | 0.941 |
| chr6 | 0.994 | 0.986 | 0.708 | 0.159 | -0.225 | 0.966 | -0.478 | 0.557 | 0.958 | -0.375 | 0.947 |
| chr7 | 0.977 | 0.987 | 0.785 | 0.280 | -0.098 | 0.958 | -0.358 | 0.648 | 0.964 | -0.253 | 0.972 |
| chr8 | 0.986 | 0.983 | 0.706 | 0.164 | -0.221 | 0.958 | -0.473 | 0.552 | 0.963 | -0.377 | 0.954 |
| chr9 | 0.905 | 0.959 | 0.897 | 0.470 | 0.104 | 0.923 | -0.158 | 0.793 | 0.939 | -0.044 | 0.967 |
| chr10 | 0.897 | 0.954 | 0.909 | 0.504 | 0.138 | 0.924 | -0.128 | 0.807 | 0.945 | -0.013 | 0.971 |
| chr11 | 0.873 | 0.940 | 0.920 | 0.538 | 0.183 | 0.900 | -0.081 | 0.825 | 0.925 | 0.028 | 0.963 |
| chr12 | 0.945 | 0.973 | 0.833 | 0.380 | 0.017 | 0.949 | -0.249 | 0.727 | 0.961 | -0.156 | 0.978 |
| chr13 | **1.000** | 0.974 | 0.663 | 0.098 | -0.283 | 0.952 | -0.527 | 0.504 | 0.940 | -0.426 | 0.925 |
| chr14 | 0.974 | **1.000** | 0.779 | 0.255 | -0.125 | 0.964 | -0.382 | 0.636 | 0.964 | -0.279 | 0.970 |
| chr15 | 0.663 | 0.779 | **1.000** | 0.763 | 0.478 | 0.731 | 0.241 | 0.936 | 0.785 | 0.358 | 0.838 |
| chr16 | 0.098 | 0.255 | 0.763 | **1.000** | 0.910 | 0.226 | 0.778 | 0.857 | 0.311 | 0.838 | 0.392 |
| chr17 | -0.283 | -0.125 | 0.478 | 0.910 | **1.000** | -0.137 | 0.954 | 0.635 | -0.057 | 0.965 | 0.024 |
| chr18 | 0.952 | 0.964 | 0.731 | 0.226 | -0.137 | **1.000** | -0.408 | 0.601 | 0.949 | -0.303 | 0.928 |
| chr19 | -0.527 | -0.382 | 0.241 | 0.778 | 0.954 | -0.408 | **1.000** | 0.429 | -0.323 | 0.977 | -0.236 |
| chr20 | 0.504 | 0.636 | 0.936 | 0.857 | 0.635 | 0.601 | 0.429 | **1.000** | 0.648 | 0.516 | 0.724 |
| chr21 | 0.940 | 0.964 | 0.785 | 0.311 | -0.057 | 0.949 | -0.323 | 0.648 | **1.000** | -0.222 | 0.957 |
| chr22 | -0.426 | -0.279 | 0.358 | 0.838 | 0.965 | -0.303 | 0.977 | 0.516 | -0.222 | **1.000** | -0.151 |
| chrX | 0.925 | 0.970 | 0.838 | 0.392 | 0.024 | 0.928 | -0.236 | 0.724 | 0.957 | -0.151 | **1.000** |

**Table 4.** Genome-wide chromosome-level correlation matrix of TFBS density changes acquired during mammalian evolution within chimpanzee brain development regulatory regions defined by the ATAC-seq analysis.

| Chromosome | chr1 | chr2 | chr3 | chr4 | chr5 | chr6 | chr7 | chr8 | chr9 | chr10 | chr11 | chr12 |
|---|---|---|---|---|---|---|---|---|---|---|---|---|
| chr1  | **1.000** | 0.706 | 0.514 | -0.142 | 0.366 | 0.265 | 0.872 | 0.674 | 0.943 | 0.726 | 0.963 | 0.851 |
| chr2  | 0.706 | **1.000** | 0.913 | 0.539 | 0.867 | 0.802 | 0.933 | 0.947 | 0.866 | 0.941 | 0.821 | 0.930 |
| chr3  | 0.514 | 0.913 | **01.00** | 0.770 | 0.968 | 0.952 | 0.850 | 0.967 | 0.715 | 0.935 | 0.674 | 0.872 |
| chr4  | -0.142 | 0.539 | 0.770 | **1.000** | 0.849 | 0.904 | 0.347 | 0.620 | 0.140 | 0.549 | 0.072 | 0.385 |
| chr5  | 0.366 | 0.867 | 0.968 | 0.849 | **1.000** | 0.966 | 0.751 | 0.907 | 0.613 | 0.873 | 0.538 | 0.776 |
| chr6  | 0.265 | 0.802 | 0.952 | 0.904 | 0.966 | **1.000** | 0.681 | 0.877 | 0.501 | 0.817 | 0.452 | 0.712 |
| chr7  | 0.872 | 0.933 | 0.850 | 0.347 | 0.751 | 0.681 | **1.000** | 0.934 | 0.959 | 0.945 | 0.943 | 0.989 |
| chr8  | 0.674 | 0.947 | 0.967 | 0.620 | 0.907 | 0.877 | 0.934 | **1.000** | 0.833 | 0.968 | 0.799 | 0.946 |
| chr9  | 0.943 | 0.866 | 0.715 | 0.140 | 0.613 | 0.501 | 0.959 | 0.833 | **1.000** | 0.882 | 0.978 | 0.947 |
| chr10 | 0.726 | 0.941 | 0.935 | 0.549 | 0.873 | 0.817 | 0.945 | 0.968 | 0.882 | **1.000** | 0.855 | 0.956 |
| chr11 | 0.963 | 0.821 | 0.674 | 0.072 | 0.538 | 0.452 | 0.943 | 0.799 | 0.978 | 0.855 | **1.000** | 0.932 |
| chr12 | 0.851 | 0.930 | 0.872 | 0.385 | 0.776 | 0.712 | 0.989 | 0.946 | 0.947 | 0.956 | 0.932 | **1.000** |
| chr13 | 0.005 | 0.642 | 0.849 | 0.976 | 0.913 | 0.953 | 0.471 | 0.721 | 0.279 | 0.661 | 0.212 | 0.508 |
| chr14 | 0.950 | 0.850 | 0.708 | 0.126 | 0.588 | 0.493 | 0.961 | 0.824 | 0.989 | 0.877 | 0.985 | 0.947 |
| chr15 | 0.986 | 0.665 | 0.453 | -0.197 | 0.316 | 0.207 | 0.839 | 0.625 | 0.926 | 0.688 | 0.943 | 0.816 |
| chr16 | 0.974 | 0.582 | 0.341 | -0.320 | 0.202 | 0.079 | 0.766 | 0.522 | 0.887 | 0.596 | 0.906 | 0.738 |
| chr17 | 0.935 | 0.452 | 0.190 | -0.467 | 0.031 | -0.072 | 0.659 | 0.389 | 0.793 | 0.456 | 0.834 | 0.628 |
| chr18 | 0.466 | 0.902 | 0.938 | 0.741 | 0.908 | 0.907 | 0.801 | 0.897 | 0.675 | 0.881 | 0.651 | 0.820 |
| chr19 | 0.909 | 0.380 | 0.134 | -0.514 | -0.035 | -0.122 | 0.610 | 0.335 | 0.745 | 0.402 | 0.800 | 0.581 |
| chr20 | 0.983 | 0.617 | 0.386 | -0.275 | 0.243 | 0.127 | 0.797 | 0.565 | 0.903 | 0.627 | 0.921 | 0.770 |
| chr21 | 0.977 | 0.659 | 0.458 | -0.188 | 0.316 | 0.215 | 0.834 | 0.622 | 0.922 | 0.677 | 0.940 | 0.813 |
| chr22 | 0.934 | 0.457 | 0.188 | -0.466 | 0.041 | -0.079 | 0.660 | 0.385 | 0.804 | 0.462 | 0.834 | 0.625 |
| chrX  | 0.917 | 0.810 | 0.710 | 0.137 | 0.561 | 0.515 | 0.927 | 0.821 | 0.926 | 0.844 | 0.954 | 0.924 |

| chr13 | chr14 | chr15 | chr16 | chr17 | chr18 | chr19 | chr20 | chr21 | chr22 | chrX |
|---|---|---|---|---|---|---|---|---|---|---|
| 0.005 | 0.950 | 0.986 | 0.974 | 0.935 | 0.466 | 0.909 | 0.983 | 0.977 | 0.934 | 0.917 |
| 0.642 | 0.850 | 0.665 | 0.582 | 0.452 | 0.902 | 0.380 | 0.617 | 0.659 | 0.457 | 0.810 |
| 0.849 | 0.708 | 0.453 | 0.341 | 0.190 | 0.938 | 0.134 | 0.386 | 0.458 | 0.188 | 0.710 |
| 0.976 | 0.126 | -0.197 | -0.320 | -0.467 | 0.741 | -0.514 | -0.275 | -0.188 | -0.466 | 0.137 |
| 0.913 | 0.588 | 0.316 | 0.202 | 0.031 | 0.908 | -0.035 | 0.243 | 0.316 | 0.041 | 0.561 |
| 0.953 | 0.493 | 0.207 | 0.079 | -0.072 | 0.907 | -0.122 | 0.127 | 0.215 | -0.079 | 0.515 |
| 0.471 | 0.961 | 0.839 | 0.766 | 0.659 | 0.801 | 0.610 | 0.797 | 0.834 | 0.660 | 0.927 |
| 0.721 | 0.824 | 0.625 | 0.522 | 0.389 | 0.897 | 0.335 | 0.565 | 0.622 | 0.385 | 0.821 |
| 0.279 | 0.989 | 0.926 | 0.887 | 0.793 | 0.675 | 0.745 | 0.903 | 0.922 | 0.804 | 0.926 |
| 0.661 | 0.877 | 0.688 | 0.596 | 0.456 | 0.881 | 0.402 | 0.627 | 0.677 | 0.462 | 0.844 |
| 0.212 | 0.985 | 0.943 | 0.906 | 0.834 | 0.651 | 0.800 | 0.921 | 0.940 | 0.834 | 0.954 |
| 0.508 | 0.947 | 0.816 | 0.738 | 0.628 | 0.820 | 0.581 | 0.770 | 0.813 | 0.625 | 0.924 |
| **1.000** | 0.263 | -0.046 | -0.172 | -0.330 | 0.820 | -0.377 | -0.128 | -0.033 | -0.331 | 0.275 |
| 0.263 | **1.000** | 0.931 | 0.890 | 0.804 | 0.670 | 0.762 | 0.907 | 0.928 | 0.811 | 0.943 |
| -0.046 | 0.931 | **1.000** | 0.983 | 0.949 | 0.427 | 0.926 | 0.986 | 0.986 | 0.951 | 0.887 |
| -0.172 | 0.890 | 0.983 | **1.000** | 0.978 | 0.314 | 0.956 | 0.994 | 0.975 | 0.984 | 0.834 |
| -0.330 | 0.804 | 0.949 | 0.978 | **1.000** | 0.180 | 0.992 | 0.972 | 0.942 | 0.994 | 0.773 |
| 0.820 | 0.670 | 0.427 | 0.314 | 0.180 | **1.000** | 0.127 | 0.355 | 0.449 | 0.169 | 0.678 |
| -0.377 | 0.762 | 0.926 | 0.956 | 0.992 | 0.127 | **1.000** | 0.949 | 0.925 | 0.981 | 0.747 |
| -0.128 | 0.907 | 0.986 | 0.994 | 0.972 | 0.355 | 0.949 | **1.000** | 0.980 | 0.974 | 0.860 |
| -0.033 | 0.928 | 0.986 | 0.975 | 0.942 | 0.449 | 0.925 | 0.980 | **1.000** | 0.939 | 0.893 |
| -0.331 | 0.811 | 0.951 | 0.984 | 0.994 | 0.169 | 0.981 | 0.974 | 0.939 | **1.000** | 0.759 |
| 0.275 | 0.943 | 0.887 | 0.834 | 0.773 | 0.678 | 0.747 | 0.860 | 0.893 | 0.759 | **1.000** |

**Figure 1.**

A

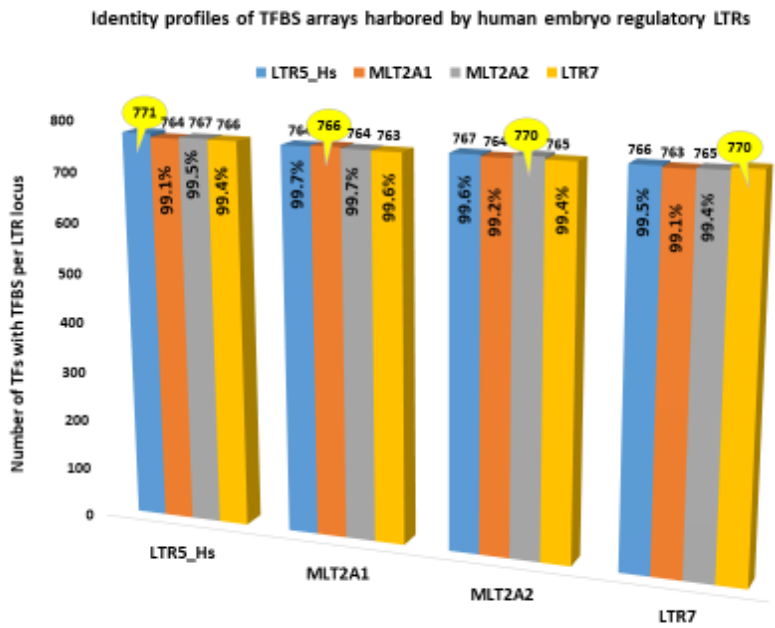

B

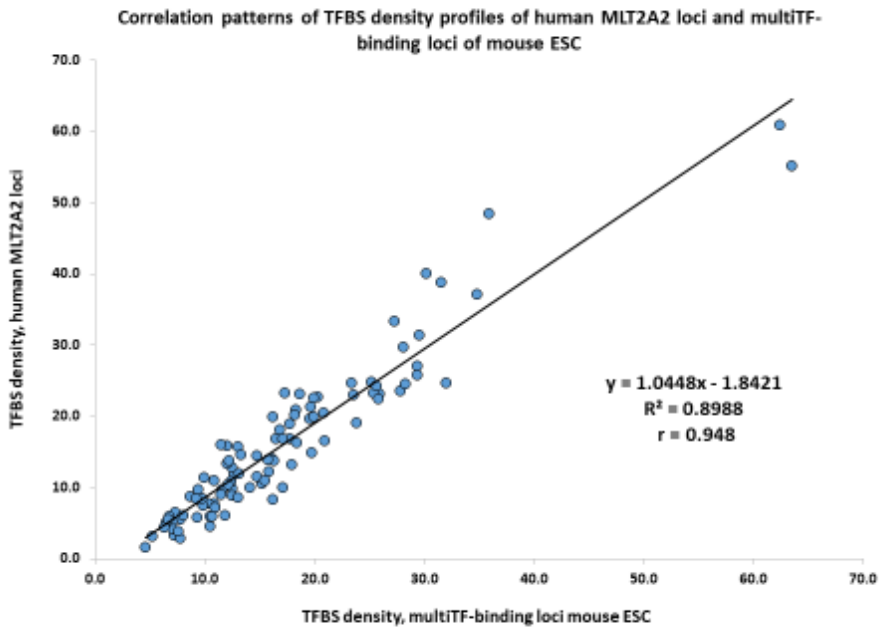

C

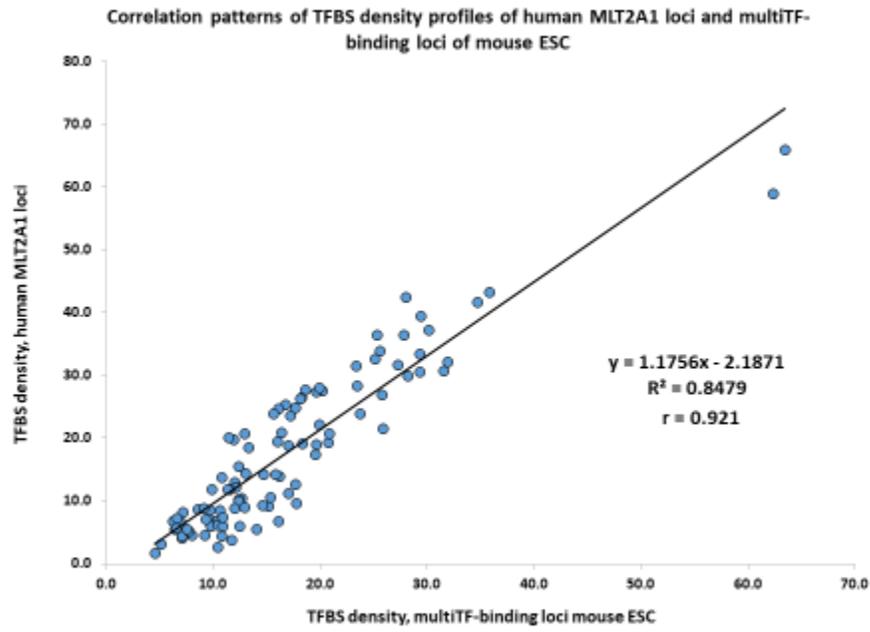

D

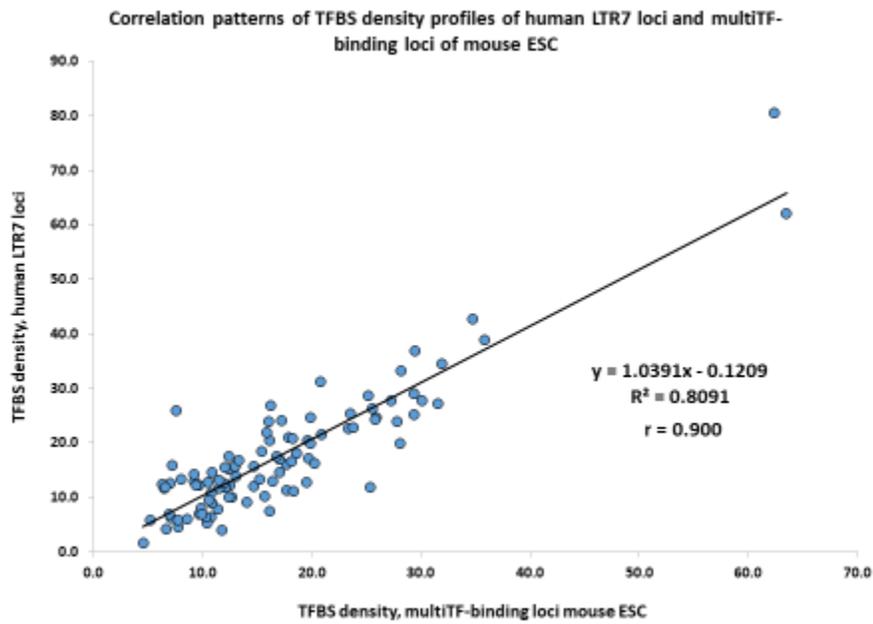

E

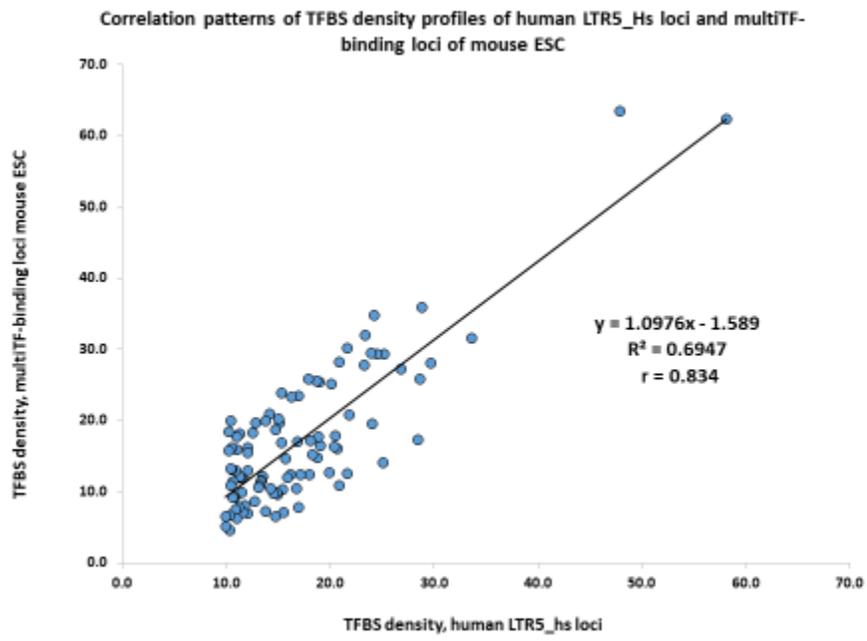

**Figure 2.**

### A. Transcription factors manifesting largest TFBS density gains compared with mouse ESC multi-TFs-binding loci

Bar chart of TFBS density gains (%) versus mouse ESC multi-TFs-binding loci, grouped by:
- LTR5_Hs regulatory LTRs: SP5, Sox6, Sox3, FOXG1, ATOH7, ATOH1, TFAP2E, Arnt, TFAP4, PRDM1, TFAP2A, KLF14, TFAP2B, KLF1, MAX, TFAP2C, CLOCK, NEUROG2, NEUROG1, SOX2
- LTR7 regulatory LTRs: ZNF354C, TBX4, NR2C1, ZEB1, Arnt, Sox3, SOX2, KLF14, KLF1, VEZF1
- MLT2A1 regulatory LTRs: BSX, FOXP3, FOXC1, EVX1, BARX1, OTX1, NKX6-3
- MLT2A2 regulatory LTRs: NKX6-3

### B. Transcription factors of human embryo LTRs manifesting more than 50% gain/loss of TFBS density

**MLT2A2 TFs: Gain: 0; Loss: 4**

| TF Name | MLT2A2 vs mouse gain |
|---|---|
| NKX6-3 | 39.5 |

| TF Name | MLT2A2 vs mouse loss |
|---|---|
| CLOCK | -53.9 |
| SOX10 | -56.2 |
| ZNF257 | -62.9 |
| NEUROG1 | -63.3 |

**MLT2A1 TFs: Gain: 7; Loss: 8**

| TF Name | MLT2A1 vs mouse gain |
|---|---|
| NKX6-3 | 74.8 |
| OTX1 | 65.1 |
| BARX1 | 60.0 |
| EVX1 | 53.4 |
| FOXC1 | 52.7 |
| FOXP3 | 50.9 |
| BSX | 50.3 |

| TF Name | MLT2A1 vs mouse loss |
|---|---|
| SNAI1 | -51.3 |
| TFAP2A | -52.0 |
| STAT3 | -57.7 |
| TFAP2C | -59.5 |
| TFAP2B | -61.3 |
| NEUROG1 | -64.5 |
| Stat5a::Stat5b | -68.2 |
| TFAP2E | -75.3 |

**LTR7 TFs: Gain: 9; Loss: 4**

| TF Name | LTR7 vs mouse gain |
|---|---|
| VEZF1 | 245.2 |
| KLF1 | 119.7 |
| KLF14 | 98.7 |
| SOX2 | 79.5 |
| Sox3 | 79.5 |
| Arnt | 76.9 |
| ZEB1 | 64.9 |
| NR2C1 | 64.6 |
| TBX4 | 55.6 |

| TF Name | LTR7 vs mouse loss |
|---|---|
| FOXL1 | -53.3 |
| STAT3 | -54.1 |
| NEUROG1 | -64.3 |
| Stat5a::Stat5b | -65.6 |

**LTR5_Hs TFs: Gain: 19; Loss: 0**

| TF Name | LTR5_Hs vs mouse gain |
|---|---|
| SOX2 | 128.0 |
| NEUROG1 | 127.3 |
| NEUROG2 | 119.4 |
| CLOCK | 117.5 |
| TFAP2C | 92.9 |
| MAX | 91.7 |
| KLF1 | 90.8 |
| TFAP2B | 78.1 |
| KLF14 | 75.1 |
| TFAP2A | 73.7 |
| PRDM1 | 73.0 |
| TFAP4 | 64.3 |
| Arnt | 63.8 |
| TFAP2E | 60.6 |
| ATOH1 | 57.1 |
| ATOH7 | 54.8 |
| FOXG1 | 54.1 |
| Sox3 | 51.4 |
| Sox6 | 50.9 |

### GSEA of 30 TFs of human embryo LTRs manifesting more than 50% gain of TFBS density

| Term | P-value | Adjusted P-value | Odds Ratio | Combined Score | Genes | Jensen TISSUES |
|---|---|---|---|---|---|---|
| Neural crest | 1.14E-09 | 2.43E-07 | 46.09127 | 949.1662 | SOX2;TFAP2A;TFAP2B;FOXC1;TFAP2C;BARX1;NEUROG1 | |
| Neural stem cell | 1.77E-08 | 1.89E-06 | 45.13636 | 805.6528 | ATOH7;SOX2;SOX3;FOXG1;NEUROG1;NEUROG2 | |
| Ectoderm | 4.59E-08 | 3.26E-06 | 38.15385 | 644.6456 | TFAP2A;SOX3;FOXG1;SOX6;NEUROG1;NEUROG2 | |
| Peripheral nerve | 8.54E-05 | 0.004549 | 19.9266 | 186.6711 | SOX2;NEUROG1;ATOH1;NEUROG2 | |

| Term | P-value | Adjusted P-value | Odds Ratio | Combined Score | Genes | PanglaoDB Augmented 2021 |
|---|---|---|---|---|---|---|
| Neural Stem/Precursor Cells | 2.90E-06 | 1.02E-04 | 26.97007 | 343.8767 | SOX2;SOX3;OTX1;NEUROG1;NEUROG2 | |
| Neuroblasts | 0.001339 | 0.023433 | 15.40559 | 101.9208 | SOX2;NEUROG1;NEUROG2 | |

| Term | P-value | Adjusted P-value | Odds Ratio | Combined Score | Genes | CellMarker Augmented 2021 |
|---|---|---|---|---|---|---|
| Germ cell:Embryo | 1.19E-04 | 0.016252 | 158.4206 | 1432.05 | TFAP2C;PRDM1 | |
| Primordial Germ cell:Undefined | 4.06E-04 | 0.02252 | 79.1746 | 618.2139 | SOX2;PRDM1 | |
| Neural Crest cell:Undefined | 4.93E-04 | 0.02252 | 71.25 | 542.5478 | NEUROG1;NEUROG2 | |

C

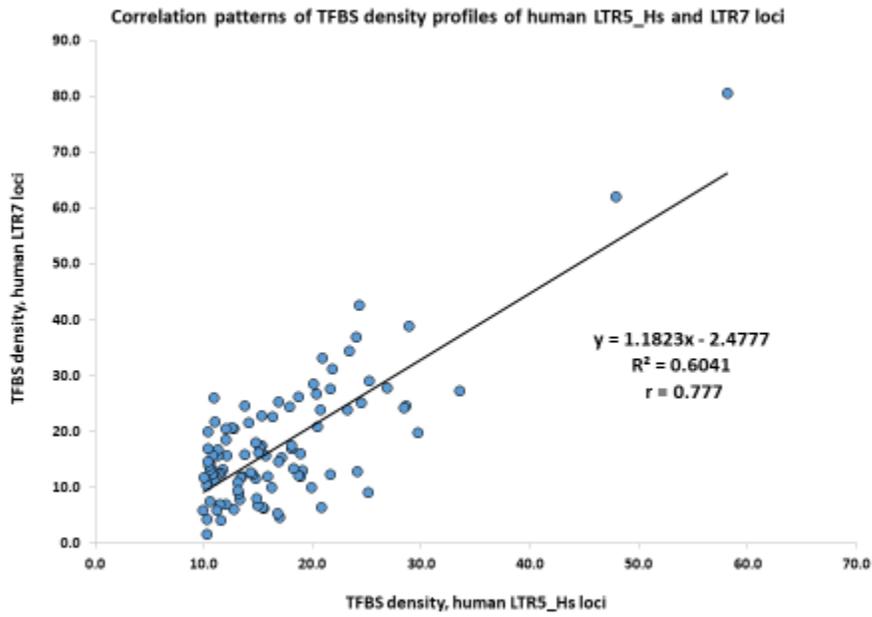

D

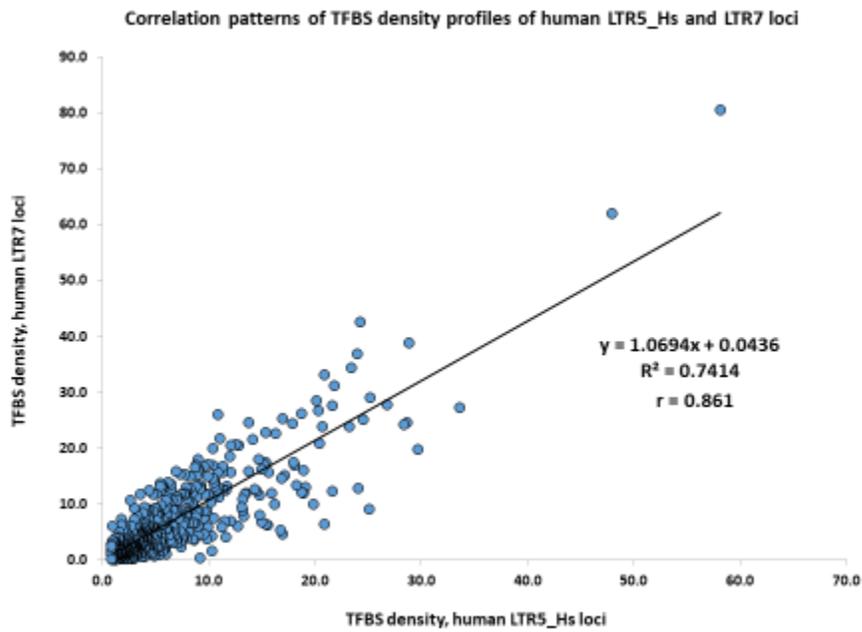

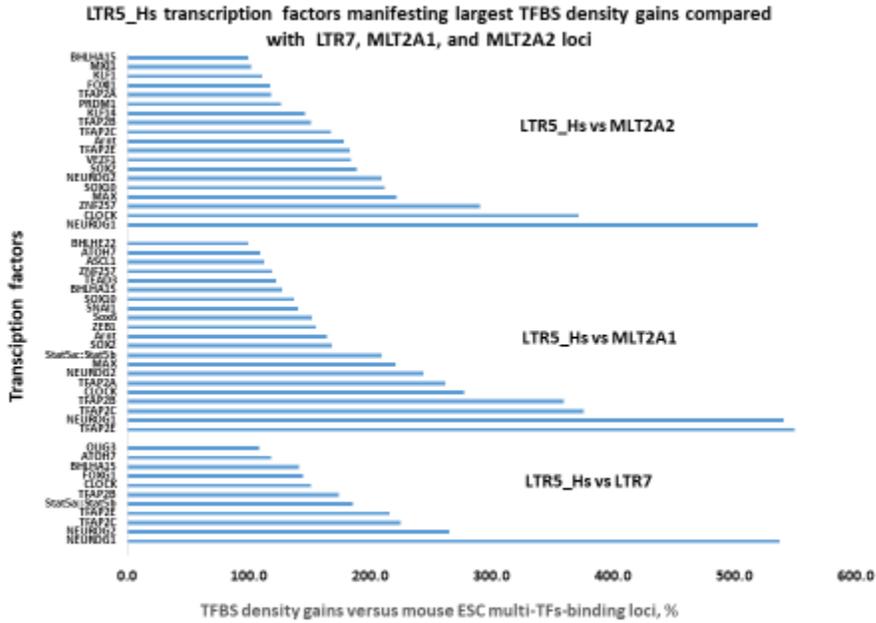

E. LTR5_Hs transcription factors manifesting largest TFBS density gains compared with LTR7, MLT2A1, and MLT2A2 loci

F.

Gain: 23 TFs; Loss: 1 TF

| TF Name | LTR5_Hs vs LTR7 gain |
|---|---|
| NEUROG1 | 537.3 |
| NEUROG2 | 265.3 |
| TFAP2C | 225.2 |
| TFAP2E | 216.5 |
| StatSa::StatSb | 186.1 |
| TFAP2B | 174.3 |
| CLOCK | 151.4 |
| FOXG1 | 144.6 |
| BHLHA15 | 141.8 |
| ATOH7 | 118.6 |
| OLIG3 | 109.0 |
| ATOH1 | 98.4 |
| ZNF257 | 89.4 |
| NFATC3 | 88.5 |
| Sox6 | 86.8 |
| TFAP2A | 76.9 |
| PRDM1 | 73.2 |
| MAX | 71.3 |
| Foxj2 | 69.3 |
| NKX2-3 | 64.7 |
| FOXO6 | 61.4 |
| FOXL1 | 59.7 |
| PTF1A | 55.6 |

| TF Name | LTR5_Hs vs LTR7 loss |
|---|---|
| VEZF1 | -58.1 |

Gain: 33 TFs; Loss: 1 TF

| TF Name | LTR5_Hs gain vs MLT2A2 |
|---|---|
| NEUROG1 | 519.9 |
| CLOCK | 371.9 |
| ZNF257 | 291.1 |
| MAX | 222.0 |
| SOX10 | 212.0 |
| NEUROG2 | 209.7 |
| SOX2 | 188.7 |
| VEZF1 | 183.9 |
| TFAP2E | 183.6 |
| Arnt | 178.2 |
| TFAP2C | 167.6 |
| TFAP2B | 151.3 |
| KLF14 | 146.1 |
| PRDM1 | 126.6 |
| TFAP2A | 118.6 |
| FOXI1 | 117.7 |
| KLF1 | 111.3 |
| MXI1 | 101.8 |
| BHLHA15 | 99.3 |
| Sox6 | 98.4 |
| ZEB1 | 92.7 |
| StatSa::StatSb | 91.2 |
| SNAI1 | 87.7 |
| ATOH7 | 85.6 |
| TEAD3 | 83.4 |
| Sox3 | 79.6 |
| FOXG1 | 73.9 |
| BHLHE22 | 73.1 |
| SP5 | 71.2 |
| SOX15 | 69.5 |
| ATOH1 | 60.0 |
| SOX4 | 57.9 |
| ASCL1 | 54.8 |

| TF Name | LTR5_Hs vs MLT2A2 loss |
|---|---|
| MIXL1 | -54.0 |

Gain: 36 TFs; Loss: 7 TF

| TF Name | LTR5_Hs gain vs MLT2A1 |
|---|---|
| TFAP2E | 549.5 |
| NEUROG1 | 540.6 |
| TFAP2C | 376.0 |
| TFAP2B | 359.7 |
| CLOCK | 277.7 |
| TFAP2A | 261.7 |
| NEUROG2 | 243.7 |
| MAX | 220.8 |
| StatSa::StatSb | 209.8 |
| SOX2 | 168.4 |
| Arnt | 164.3 |
| ZEB1 | 155.7 |
| Sox6 | 151.8 |
| SNAI1 | 141.0 |
| SOX10 | 137.1 |
| BHLHA15 | 127.9 |
| TEAD3 | 122.6 |
| ZNF257 | 119.4 |
| ASCL1 | 112.9 |
| ATOH7 | 109.8 |
| BHLHE22 | 99.6 |
| VEZF1 | 97.9 |
| ATOH1 | 91.1 |
| Sox3 | 78.5 |
| PRDM1 | 77.0 |
| MXI1 | 76.0 |
| SOX4 | 71.8 |
| SP5 | 71.6 |
| MEIS3 | 68.3 |
| KLF1 | 68.1 |
| KLF14 | 62.8 |
| FOXI1 | 54.6 |
| STAT3 | 54.2 |
| FIGLA | 50.9 |
| SOX15 | 50.8 |
| OLIG2 | 50.1 |

| TF Name | LTR5_Hs vs MLT2A1 loss |
|---|---|
| EVX1 | -51.2 |
| EVX2 | -52.6 |
| DRGX | -53.4 |
| Arid3a | -55.8 |
| ESX1 | -56.8 |
| FOXC1 | -57.2 |
| MIXL1 | -62.8 |

## G

**GSEA of 44 LTR5_Hs TFs manifesting more than 50% gain of TFBS density**

| Term | P-value | Adjusted P-value | Odds Ratio | Combined Score | Genes | PanglaoDB Augmented 2021 |
|---|---|---|---|---|---|---|
| Neural Stem/Precursor Cells | 9.71E-07 | 5.63E-05 | 21.42304 | 296.6198 | SOX2;SOX3;ASCL1;MEIS3;NEUROG1;NEUROG2 | |
| Neuroblasts | 2.95E-04 | 0.00855 | 13.95382 | 113.4297 | SOX2;ASCL1;NEUROG1;NEUROG2 | |
| Embryonic Stem Cells | 5.08E-04 | 0.011557 | 11.50233 | 85.37396 | SOX2;SOX15;STAT3;TEAD3 | |
| Trophoblast Stem Cells | 0.001345 | 0.019499 | 15.1372 | 100.0799 | TFAP2A;TFAP2C;TEAD3 | |
| Motor Neurons | 0.001867 | 0.02166 | 13.44715 | 84.09255 | TFAP2A;FOXO6;MEIS3 | |
| Oligodendrocyte Progenitor Cells | 0.002618 | 0.025309 | 11.80564 | 70.72271 | OLIG2;ASCL1;SOX10 | |

| Term | P-value | Adjusted P-value | Odds Ratio | Combined Score | Genes | CellMarker Augmented 2021 |
|---|---|---|---|---|---|---|
| Neural Crest cell:Undefined | 1.04E-09 | 1.74E-07 | 150.3695 | 3109.743 | SNAI1;ASCL1;SOX10;NEUROG1;NEUROG2 | |
| Oligodendrocyte:Undefined | 3.46E-07 | 2.89E-05 | 364.9756 | 5430.168 | OLIG3;OLIG2;SOX10 | |
| Brush Cell (Tuft cell):Lung | 2.45E-05 | 0.001024 | 27.23609 | 289.1436 | NFATC3;PRDM1;ASCL1;SOX4 | |
| Neuroendocrine cell:Lung | 2.45E-05 | 0.001024 | 27.23609 | 289.1436 | NFATC3;PRDM1;ASCL1;SOX4 | |
| Cancer Stem cell:Brain | 3.87E-05 | 0.001292 | 54.00813 | 548.7162 | SOX2;ASCL1;SOX4 | |
| Oligodendrocyte Progenitor cell:Embryonic Prefrontal Cortex | 9.93E-05 | 0.002762 | 18.72642 | 172.6175 | OLIG2;SOX6;SOX10;SOX4 | |
| Germ cell:Embryo | 2.57E-04 | 0.006129 | 105.5397 | 872.4795 | TFAP2C;PRDM1 | |
| Primordial Germ cell:Undefined | 8.76E-04 | 0.018294 | 52.74603 | 371.3176 | SOX2;PRDM1 | |
| Neural Stem cell:Brain | 0.001374 | 0.025407 | 41.26915 | 271.9617 | SOX2;ASCL1 | |
| Germ cell:Undefined | 0.001603 | 0.026774 | 37.96381 | 244.3252 | TFAP2C;PRDM1 | |
| Cancer Stem cell:Undefined | 0.002618 | 0.039547 | 11.80564 | 70.72271 | SOX2;STAT3;ASCL1 | |
| Induced Pluripotent Stem cell:Undefined | 0.002842 | 0.039547 | 27.90196 | 163.5993 | SOX2;STAT3 | |

| Term | P-value | Adjusted P-value | Odds Ratio | Combined Score | Genes | ARCHS4 Tissues |
|---|---|---|---|---|---|---|
| FETAL BRAIN | 6.87E-07 | 7.35E-05 | 5.32 | 75.40 | TFAP2C;FOXG1;FOXO6;BHLHE22;BHLHA15;OLIG2;ASCL1;MEIS3;ATOH7;SOX2;SOX3;ZEB1;SOX15;SP5;SOX6;NEUROG1;NEUROG2;TEAD3 | |
| CEREBELLUM | 7.44E-05 | 0.00199 | 3.97 | 37.73 | TFAP2A;TFAP2B;OLIG3;FOXO6;BHLHE22;OLIG2;ASCL1;SOX10;SOX2;SOX3;ZEB1;SP5;PTF1A;NEUROG1;ATOH1 | |
| OLIGODENDROCYTE | 7.44E-05 | 0.00199 | 3.97 | 37.73 | TFAP2B;TFAP2C;FOXG1;FOXO6;OLIG2;ASCL1;MEIS3;SOX10;ATOH7;SOX2;SOX3;SP5;SOX6;NEUROG1;TEAD3 | |
| PREFRONTAL CORTEX | 7.44E-05 | 0.00199 | 3.97 | 37.73 | TFAP2C;FOXG1;BHLHE22;FOXO6;OLIG2;ASCL1;MEIS3;SOX10;ATOH7;SOX2;ZEB1;SOX6;PTF1A;NEUROG1;SOX4;NEUROG2 | |
| SPINAL CORD | 2.94E-04 | 0.005244 | 3.58 | 29.10 | TFAP2A;TFAP2B;OLIG3;FOXO6;BHLHE22;OLIG2;ASCL1;MEIS3;SOX10;SOX2;SOX3;PTF1A;NEUROG1;NEUROG2 | |
| AMNIOTIC FLUID | 0.003421 | 0.036608 | 2.87 | 16.31 | SOX2;TFAP2A;TFAP2B;TFAP2C;SOX15;FOXI1;FOXJ2;FOXO6;SP5;PRDM1;SOX10;TEAD3 | |
| BRAIN (BULK) | 0.003421 | 0.036608 | 2.87 | 16.31 | ATOH7;SOX2;TFAP2B;TFAP2E;FOXG1;SOX15;BHLHE22;OLIG2;ASCL1;MEIS3;PTF1A;SOX10 | |
| MOTOR NEURON | 0.003421 | 0.036608 | 2.87 | 16.31 | SOX2;SOX3;TFAP2B;OLIG3;FOXO6;BHLHE22;OLIG2;ASCL1;MEIS3;SOX10;NEUROG1;NEUROG2 | |
| SUPERIOR FRONTAL GYRUS | 0.003421 | 0.036608 | 2.87 | 16.31 | ATOH7;SOX2;TFAP2E;FOXG1;SOX15;FOXO6;BHLHE22;OLIG2;ASCL1;MEIS3;PTF1A;SOX10 | |

## H

**GSEA of 44 LTR5_Hs TFs manifesting more than 50% gain of TFBS density**

| Term | P-value | Adjusted P-value | Odds Ratio | Combined Score | Genes | CellMarker 2024 |
|---|---|---|---|---|---|---|
| Progenitor Cell Brain Human | 6.09E-08 | 1.08E-05 | 142.44 | 2366.46 | SOX2;FOXG1;OLIG2;SOX10 | |
| Neuron Brain Human | 9.02E-08 | 1.08E-05 | 22.09 | 368.09 | SOX2;FOXG1;OLIG2;ASCL1;SOX6;SOX10;SOX4 | |
| Cancer Stem Cell Brain Human | 1.76E-06 | 1.29E-04 | 55.33 | 733.27 | SOX2;OLIG2;ASCL1;SOX4 | |
| Oligodendrocyte Precursor Cell Brain Human | 2.16E-06 | 1.29E-04 | 162.17 | 2115.93 | OLIG3;SOX6;SOX10 | |
| Germ Cell Ovary Mouse | 2.80E-06 | 1.34E-04 | 145.95 | 1866.18 | FIGLA;STAT3;PRDM1 | |
| Stem Cell Brain Mouse | 4.44E-06 | 1.77E-04 | 121.61 | 1498.90 | OLIG2;ASCL1;SOX10 | |
| Neural Progenitor Cell Brain Human | 6.61E-06 | 2.27E-04 | 104.23 | 1243.08 | SOX2;OLIG2;SOX4 | |
| Brush Cell (Tuft Cell) Lung Human | 2.33E-05 | 6.88E-04 | 27.02 | 294.60 | NFATC3;PRDM1;ASCL1;SOX4 | |
| Neuroendocrine Cell Lung Human | 2.58E-05 | 6.88E-04 | 26.87 | 283.85 | NFATC3;PRDM1;ASCL1;SOX4 | |
| Oligodendrocyte Precursor Cell Brain Mouse | 3.30E-05 | 7.92E-04 | 25.16 | 259.65 | ZEB1;OLIG2;SOX6;SOX10 | |
| Amacrine Cell Retina Mouse | 4.71E-05 | 9.42E-04 | 316.71 | 3155.49 | TFAP2A;TFAP2B | |
| Cranial Neural Crest Cell Skin Human | 4.71E-05 | 9.42E-04 | 316.71 | 3155.49 | TFAP2A;SOX10 | |
| Neuron Brain Mouse | 5.15E-05 | 9.51E-04 | 7.07 | 69.83 | FOXG1;VEZF1;OLIG2;ASCL1;PTF1A;SOX10;ATOH1;SOX4 | |
| Epithelial-mesenchymal Cell Bladder Human | 7.06E-05 | 0.001129 | 237.52 | 2270.52 | ZEB1;SNAI1 | |
| Globose Basal Cell Olfactory Neuroepithelium Human | 7.06E-05 | 0.001129 | 237.52 | 2270.52 | SOX2;NEUROG1 | |
| Oligodendrocyte Cortex Human | 9.86E-05 | 0.001254 | 190.01 | 1752.66 | OLIG2;SOX10 | |
| Dermal Condensate Branch Cell Hair Follicle Mouse | 9.86E-05 | 0.001254 | 190.01 | 1752.66 | SOX2;PRDM1 | |
| Mesenchymal Cell Breast Mouse | 9.86E-05 | 0.001254 | 190.01 | 1752.66 | ZEB1;SNAI1 | |
| Oligodendrocyte Progenitor Cell Embryonic Prefrontal Cortex Human | 9.93E-05 | 0.001254 | 18.73 | 172.62 | OLIG2;SOX6;SOX10;SOX4 | |
| Mesenchymal Cell Pancreas Human | 1.69E-04 | 0.002023 | 135.71 | 1179.01 | ZEB1;SNAI1 | |
| Germ Cell Embryo Human | 2.10E-04 | 0.002196 | 118.74 | 1005.25 | TFAP2C;PRDM1 | |
| Oligodendrocyte Progenitor Cell Brain Human | 2.10E-04 | 0.002196 | 118.74 | 1005.25 | OLIG2;SOX10 | |
| Embryonic Stem Cell Blood Mouse | 2.10E-04 | 0.002196 | 118.74 | 1005.25 | SOX2;STAT3 | |
| Primordial Germ Cell Embryo Mouse | 2.57E-04 | 0.002569 | 105.54 | 872.48 | SOX2;PRDM1 | |
| Basal Cell Bladder Human | 3.08E-04 | 0.002955 | 94.98 | 768.01 | TFAP2C;STAT3 | |
| Neural Stem Cell Brain Mouse | 3.63E-04 | 0.003114 | 86.34 | 683.85 | SOX2;ASCL1 | |
| Mesenchymal Cell Kidney Human | 3.63E-04 | 0.003114 | 86.34 | 683.85 | ZEB1;SNAI1 | |
| Stem Cell Bone Marrow Human | 3.63E-04 | 0.003114 | 86.34 | 683.85 | SOX2;SOX4 | |
| Epithelial Cell Kidney Human | 4.88E-04 | 0.003902 | 73.05 | 557.07 | ZEB1;SNAI1 | |
| Primordial Germ Cell Undefined Human | 4.88E-04 | 0.003902 | 73.05 | 557.07 | SOX2;PRDM1 | |
| Neuron Embryo Mouse | 7.08E-04 | 0.005308 | 59.35 | 430.46 | NEUROG1;NEUROG2 | |

I    **GSEA of 44 LTR5_Hs TFs manifesting more than 50% gain of TFBS density**

| Term | P-value | Adjusted P-value | Odds Ratio | Combined Score | Genes | Jensen TISSUES |
|---|---|---|---|---|---|---|
| Neural stem cell | 6.10E-14 | 1.44E-11 | 55.08 | 1675.93 | ATOH7;SOX2;SOX3;FOXG1;STAT3;OLIG2;ASCL1;SOX10;NEUROG1;NEUROG2 | |
| Ectoderm | 1.38E-11 | 1.63E-09 | 40.15 | 1004.03 | TFAP2A;SOX3;FOXG1;FOXI1;SNAI1;SOX6;SOX10;NEUROG1;NEUROG2 | |
| Neural crest | 6.02E-10 | 4.75E-08 | 33.89 | 719.53 | SOX2;TFAP2A;TFAP2B;TFAP2C;SNAI1;ASCL1;SOX10;NEUROG1 | |
| Neural tube | 2.29E-07 | 1.36E-05 | 27.73 | 423.89 | FOXG1;OLIG2;ASCL1;SOX10;NEUROG1;NEUROG2 | |
| Neuroblast | 3.69E-06 | 1.75E-04 | 45.25 | 566.09 | SOX2;OLIG2;ASCL1;NEUROG2 | |
| Gill arch | 9.39E-06 | 3.71E-04 | 91.19 | 1055.57 | FOXI1;NEUROG1;NEUROG2 | |
| Peripheral nerve | 2.37E-05 | 7.65E-04 | 16.70 | 177.88 | SOX2;SOX10;NEUROG1;ATOH1;NEUROG2 | |
| Cancer stem cell | 2.58E-05 | 7.65E-04 | 26.87 | 283.85 | SOX2;ZEB1;STAT3;SNAI1 | |
| Periosteum | 3.87E-05 | 0.001014 | 54.01 | 508.72 | SOX2;STAT3;BHLHE22 | |
| P-19 cell | 4.28E-05 | 0.001014 | 52.08 | 523.87 | SOX2;ASCL1;NEUROG1 | |
| Primordium | 7.98E-05 | 0.00172 | 19.86 | 187.36 | SOX2;TFAP2C;FIGLA;PRDM1 | |
| NT2/D1 cell | 1.31E-04 | 0.002594 | 158.33 | 1415.15 | SOX3;SOX15 | |
| Blastocyst | 1.84E-04 | 0.003347 | 15.86 | 136.48 | SOX2;TFAP2C;STAT3;PRDM1 | |
| Mesenchyme | 2.75E-04 | 0.004657 | 9.75 | 79.93 | ZEB1;STAT3;SNAI1;FOXL1;SOX10 | |
| Trophoblast | 4.27E-04 | 0.00646 | 12.61 | 97.83 | SOX2;TFAP2A;TFAP2C;STAT3 | |
| Colorectal cancer cell | 4.51E-04 | 0.00646 | 22.39 | 172.53 | ZEB1;STAT3;SNAI1 | |
| Gbm cell | 4.88E-04 | 0.00646 | 73.05 | 557.07 | SOX2;STAT3 | |
| Pancreatic cancer cell | 4.91E-04 | 0.00646 | 21.72 | 165.51 | STAT3;SNAI1;PTF1A | |
| Embryonic carcinoma cell | 5.57E-04 | 0.006943 | 67.83 | 508.30 | SOX2;ASCL1 | |
| Adult | 7.77E-04 | 0.009213 | 4.65 | 33.30 | SOX2;STAT3;OLIG2;ASCL1;SOX6;SOX10;NEUROG2;KLF1 | |
| Hair cell | 0.001054 | 0.0119 | 16.52 | 113.24 | SOX2;NEUROG1;ATOH1 | |
| Germ | 0.001158 | 0.011974 | 15.97 | 107.99 | SOX2;FIGLA;PRDM1 | |
| Neural plate | 0.001162 | 0.011974 | 45.20 | 305.47 | SP5;MEIS3 | |
| Breast cancer cell line | 0.001537 | 0.015178 | 8.85 | 57.32 | TFAP2C;ZEB1;STAT3;SNAI1 | |
| Multiple myeloma cell | 0.001979 | 0.018036 | 33.89 | 210.98 | STAT3;PRDM1 | |
| Supporting cell | 0.001979 | 0.018036 | 33.89 | 210.98 | SOX2;ATOH1 | |
| Finger | 0.002622 | 0.022748 | 5.79 | 34.44 | ZEB1;SNAI1;VEZF1;PRDM1;KLF1 | |
| Secretory cell | 0.002688 | 0.022748 | 28.75 | 170.17 | BHLHA15;ATOH1 | |
| Colorectum | 0.003328 | 0.027201 | 25.64 | 146.26 | ZEB1;NEUROG1 | |
| Yolk sac | 0.004034 | 0.031865 | 23.13 | 127.52 | TFAP2C;KLF1 | |
| Mammary gland cell line | 0.00441 | 0.033714 | 22.05 | 119.61 | STAT3;SNAI1 | |
| Optic nerve | 0.006991 | 0.051781 | 17.23 | 85.5151 | ATOH7;SOX3 | |

**Figure 3.**

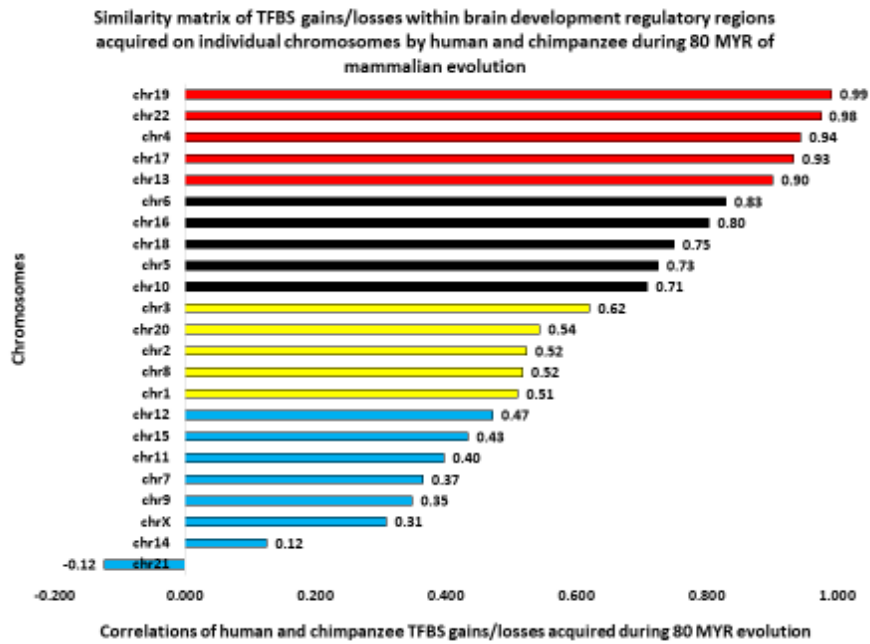

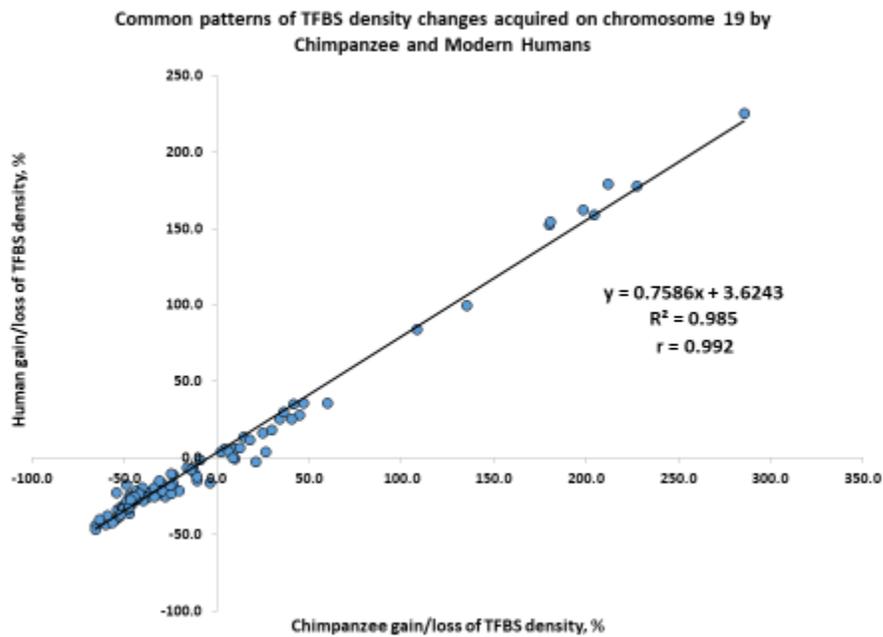

C

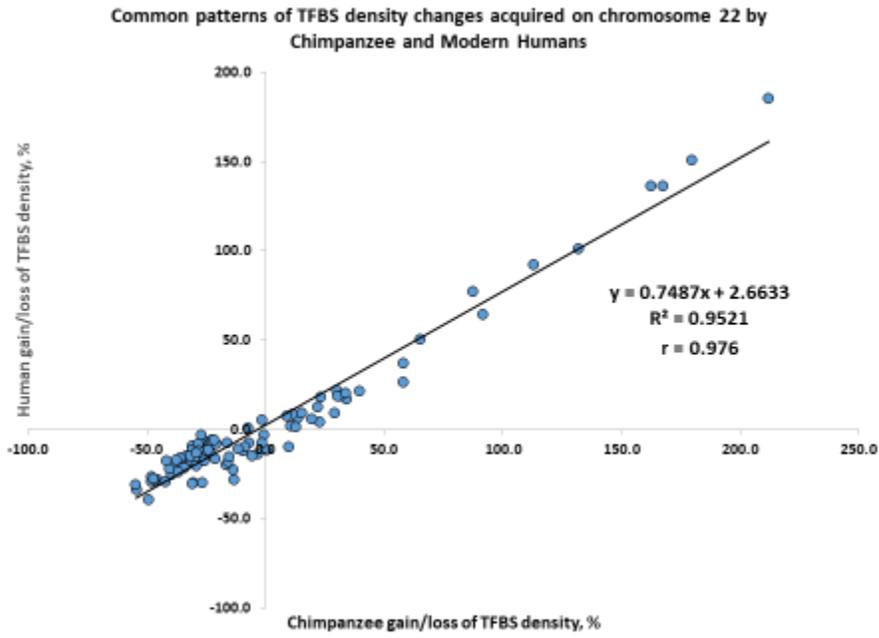

D

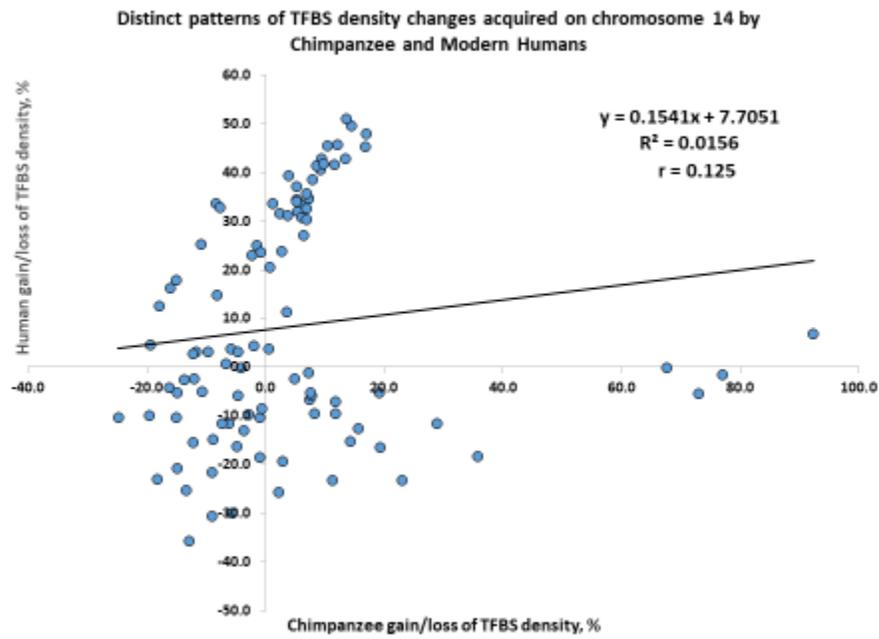

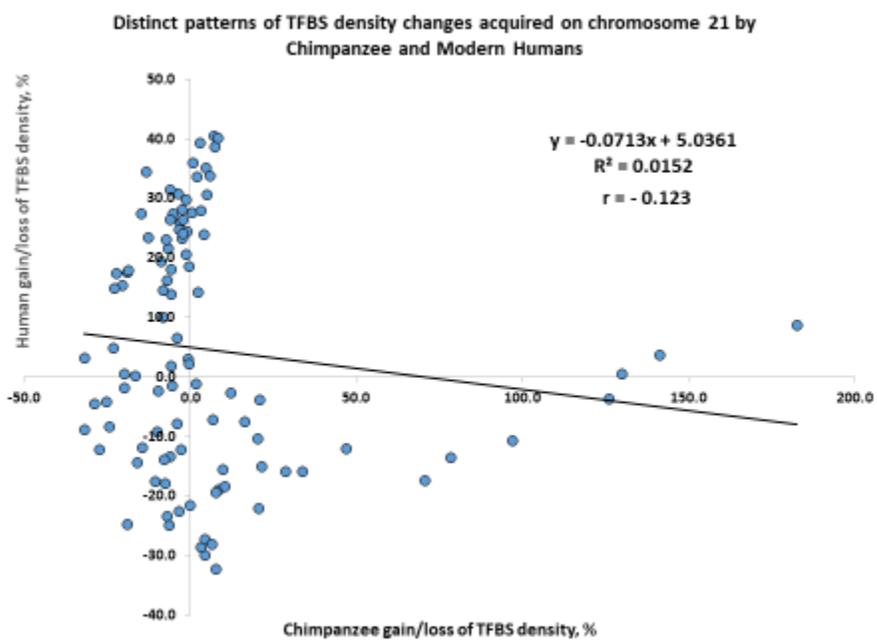

E

**Figure 4.**

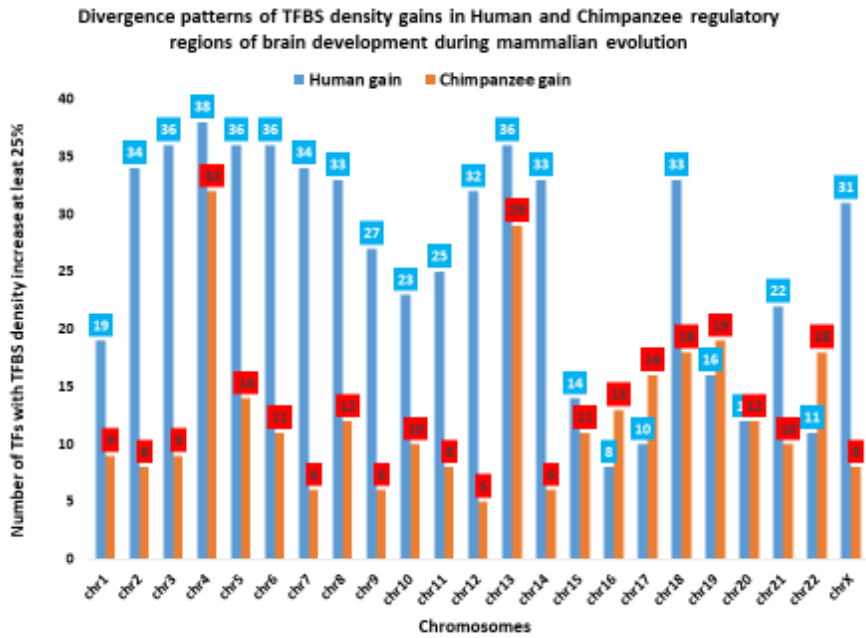

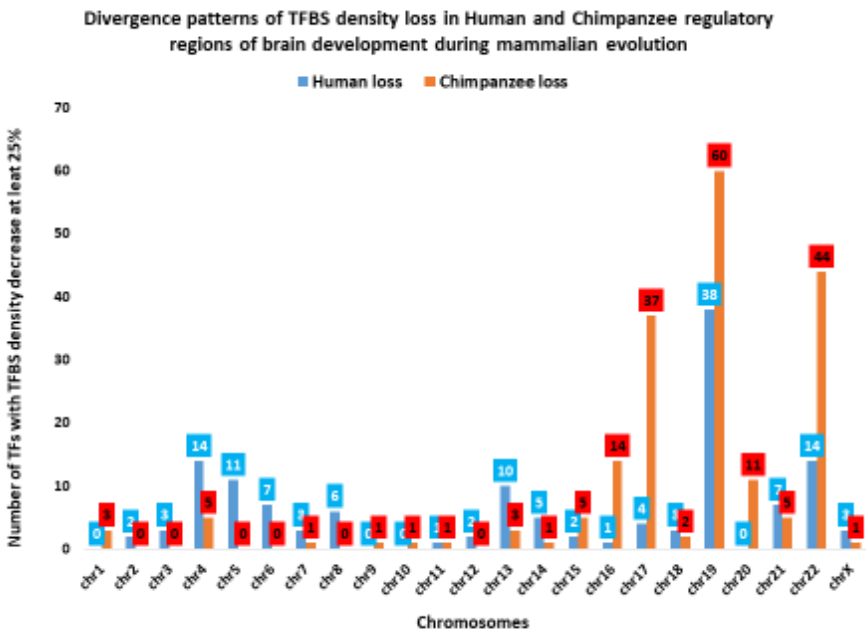

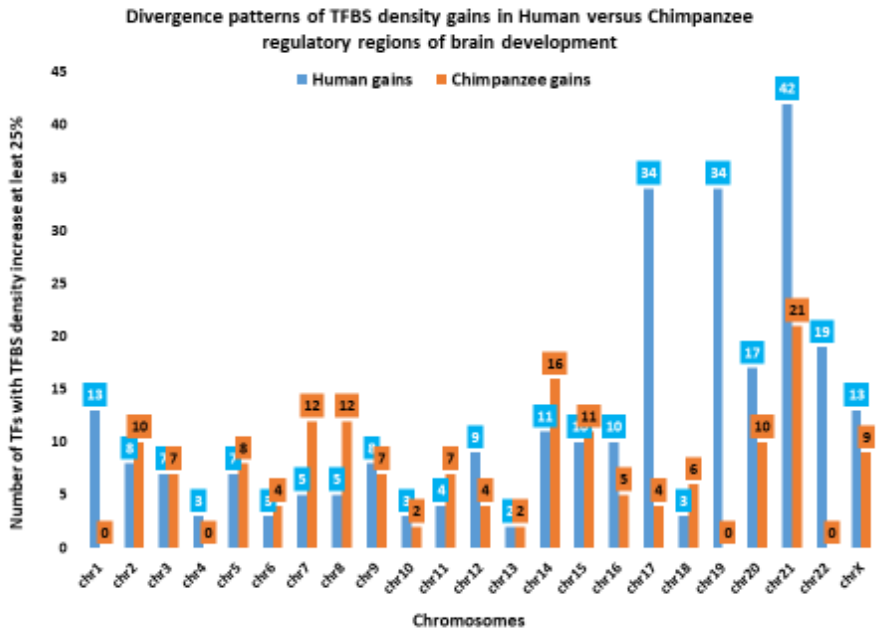

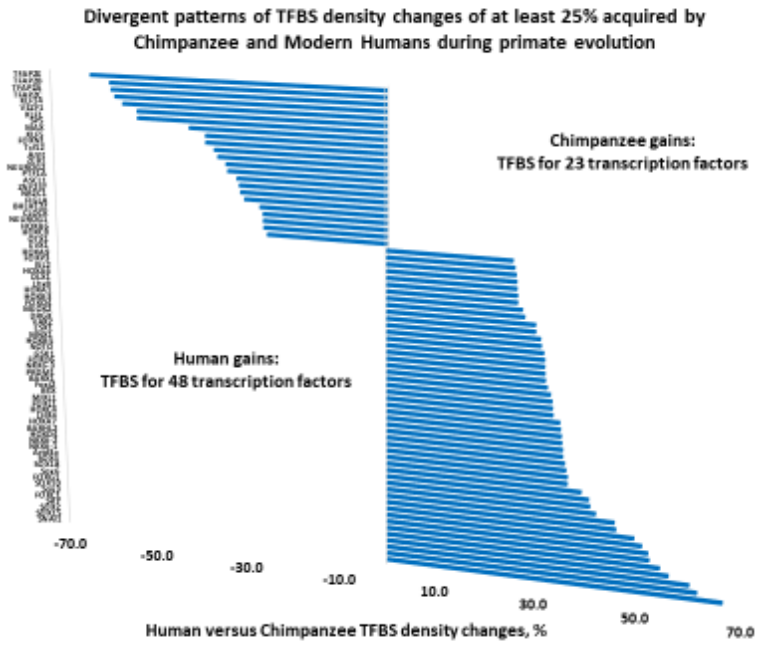

**Figure 5.**

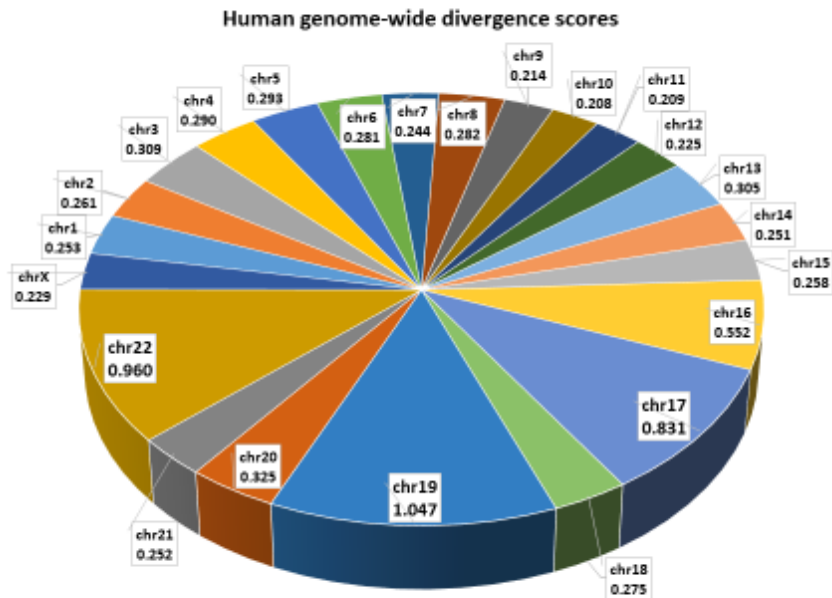

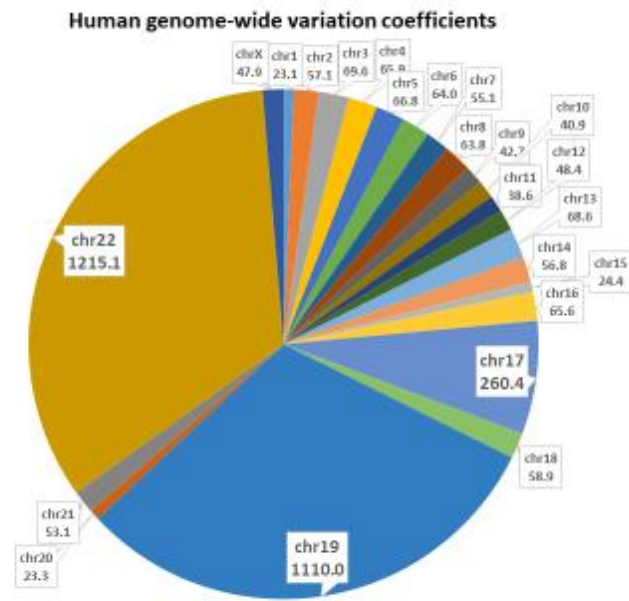

C

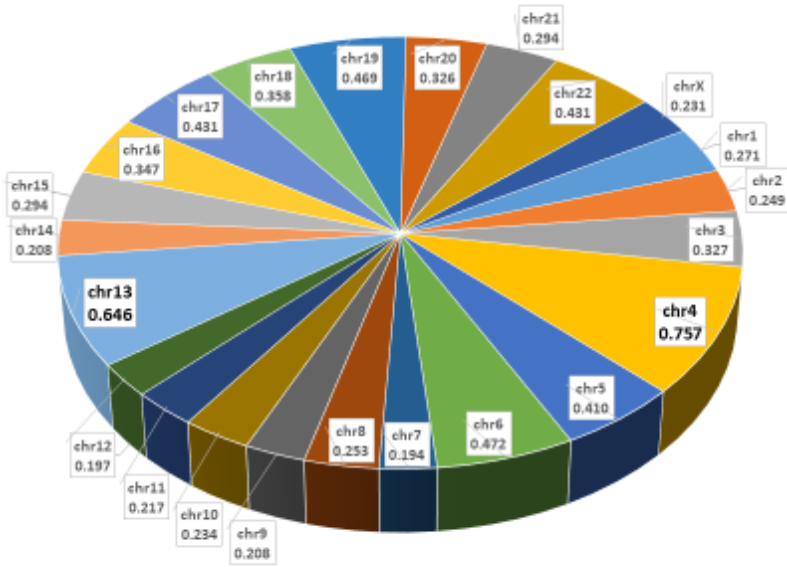

D

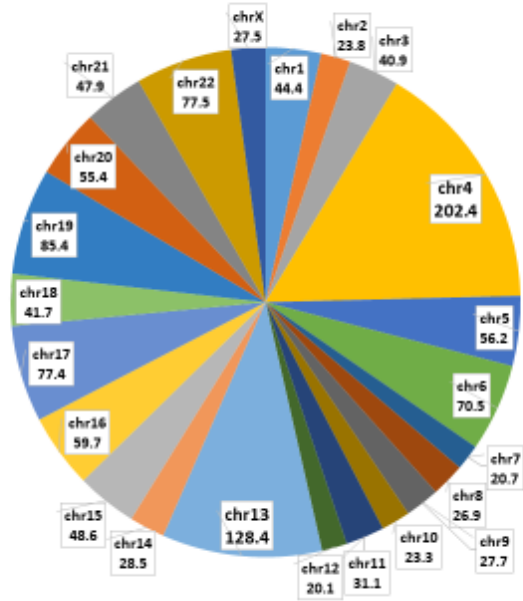

**Figure 6.**

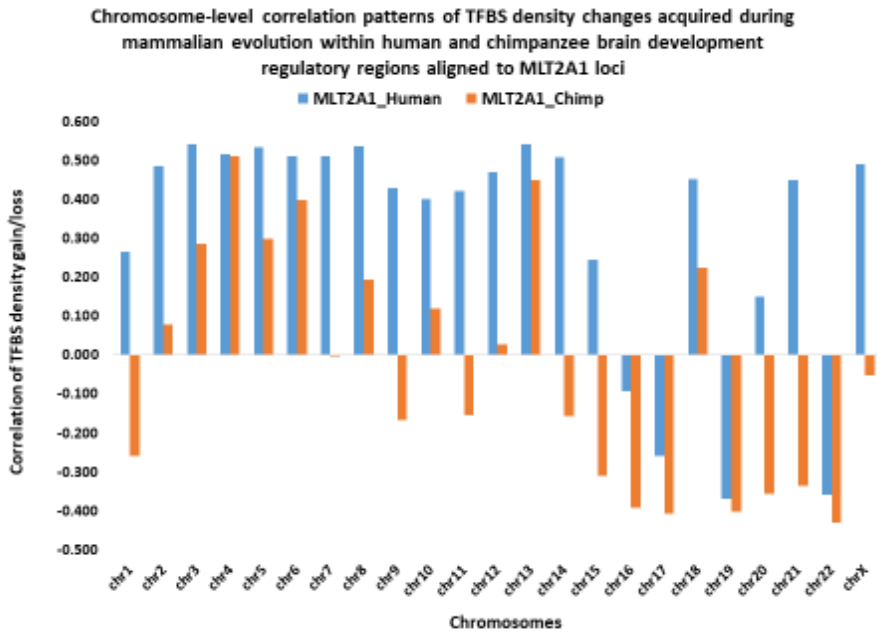

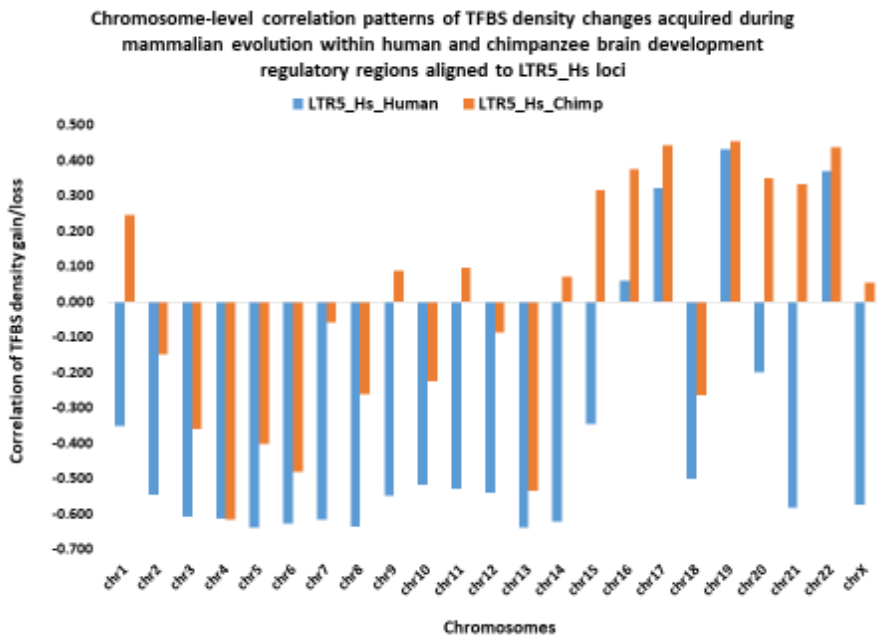

C

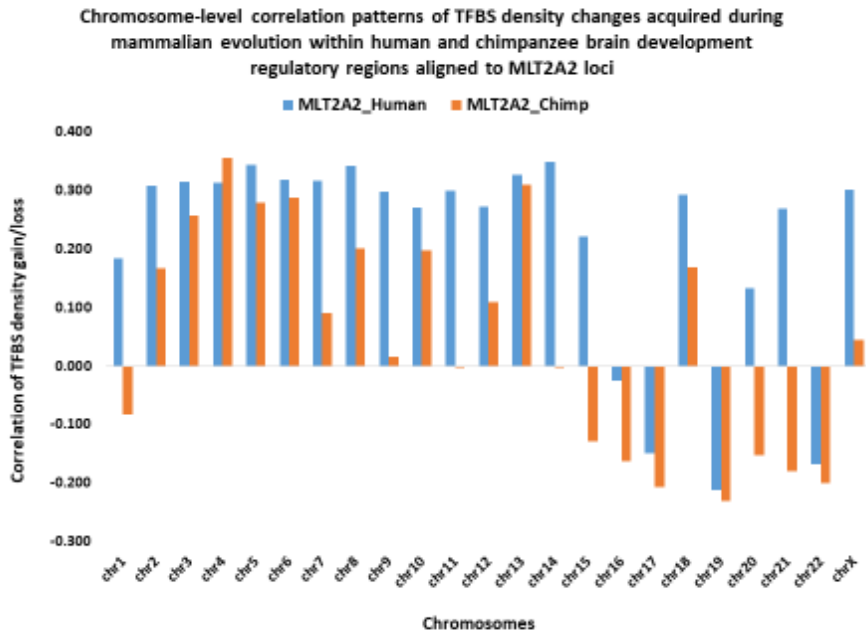

D

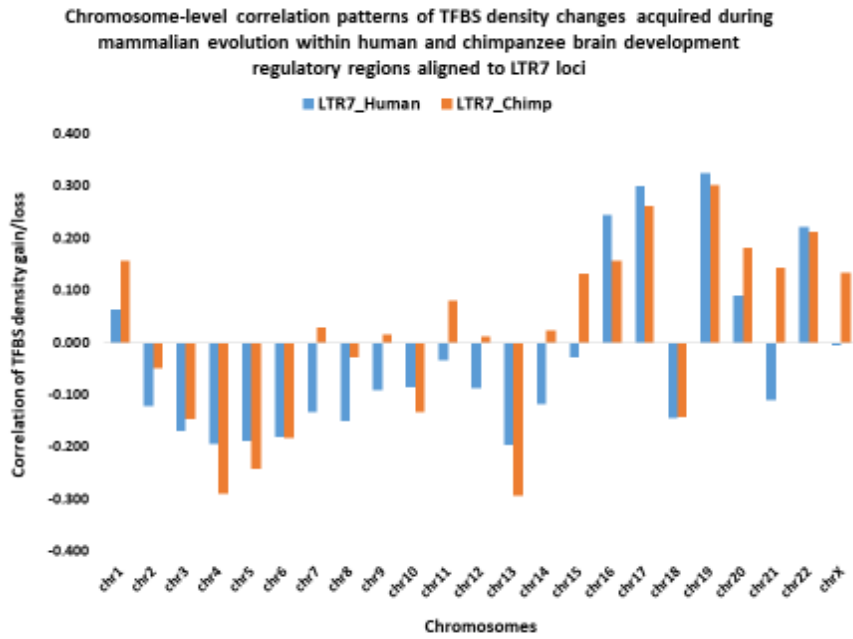

E

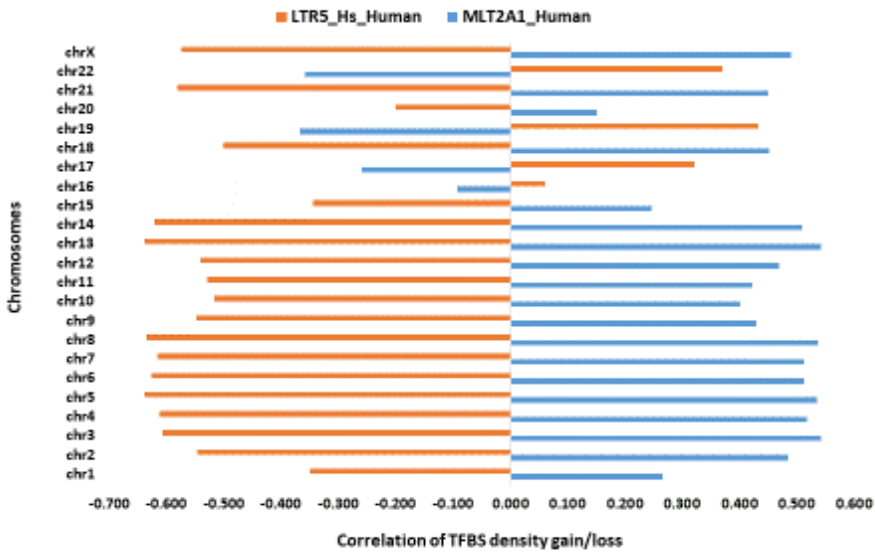

F

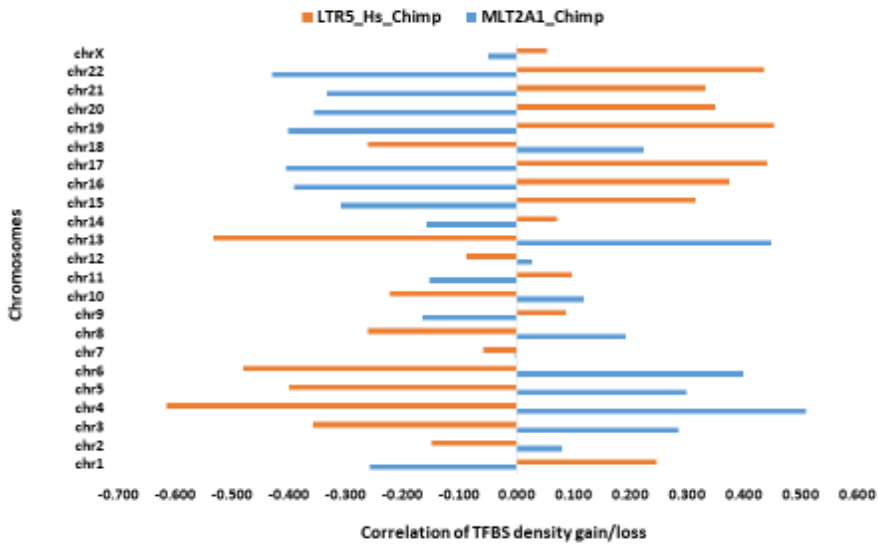

G

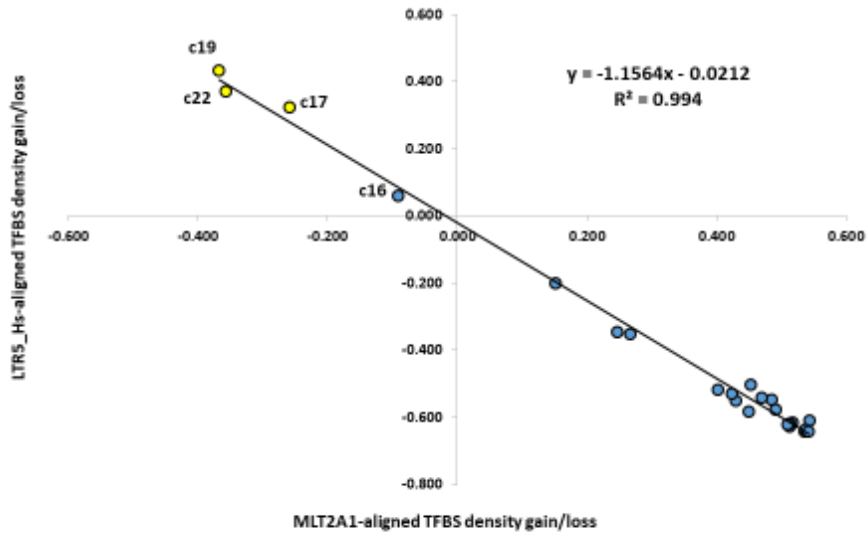

H

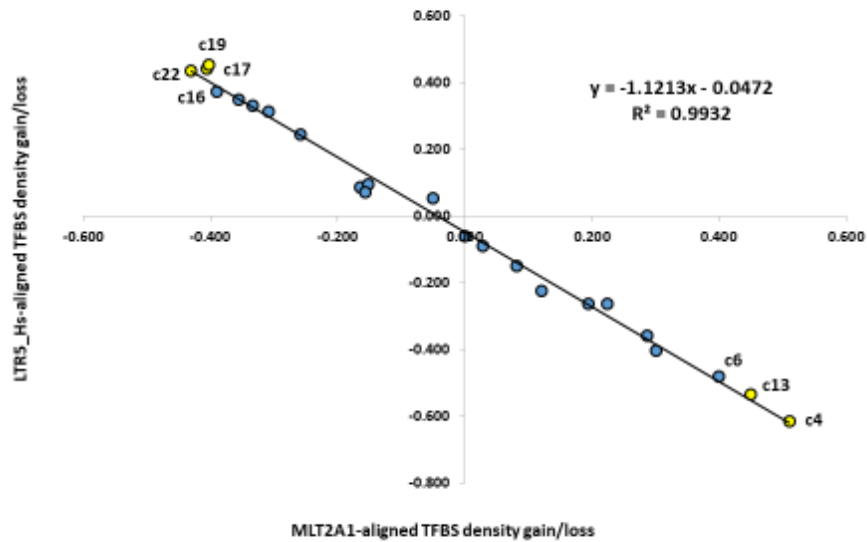

I
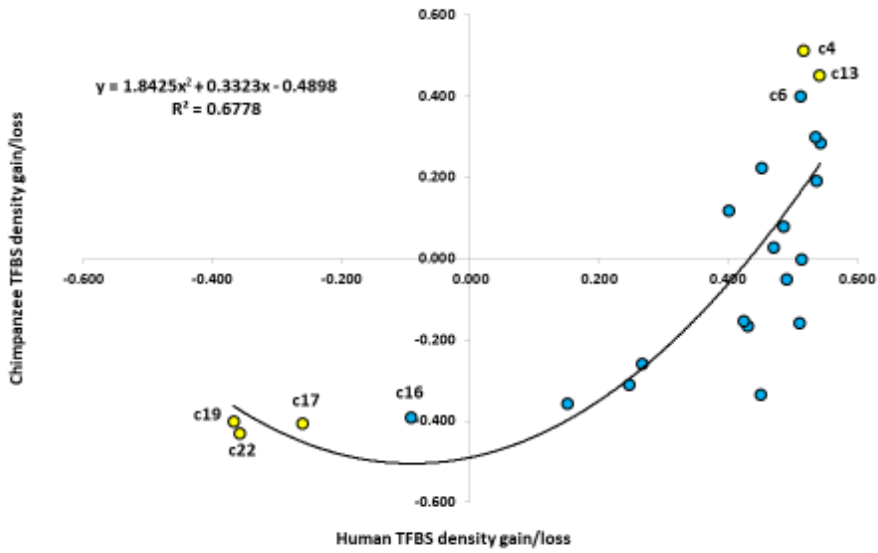

J
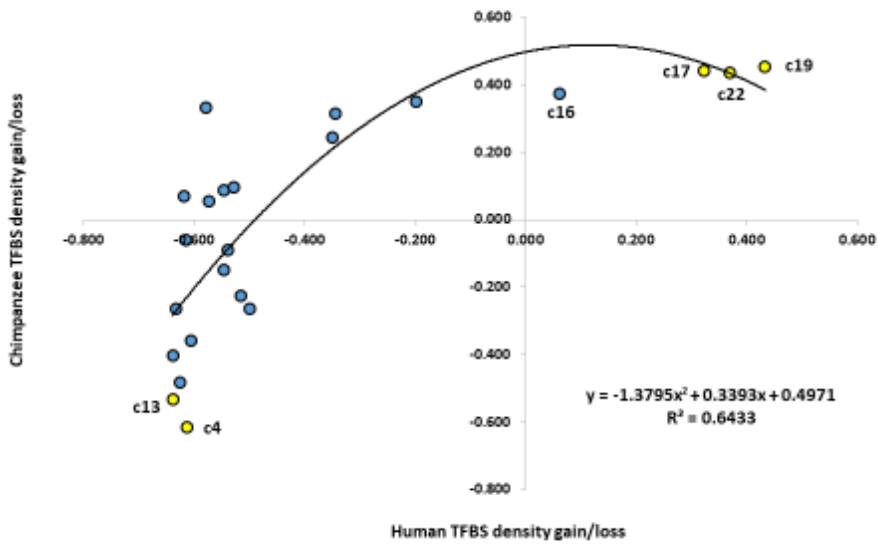

K

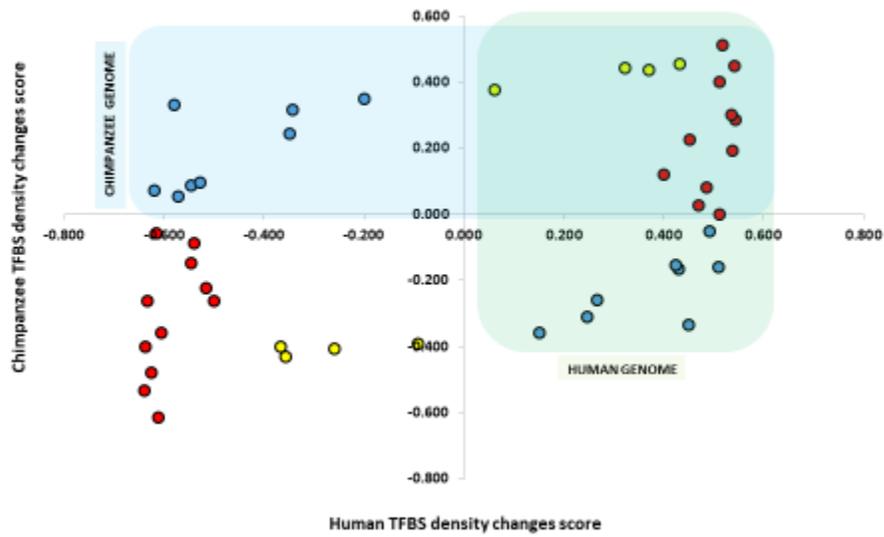

L

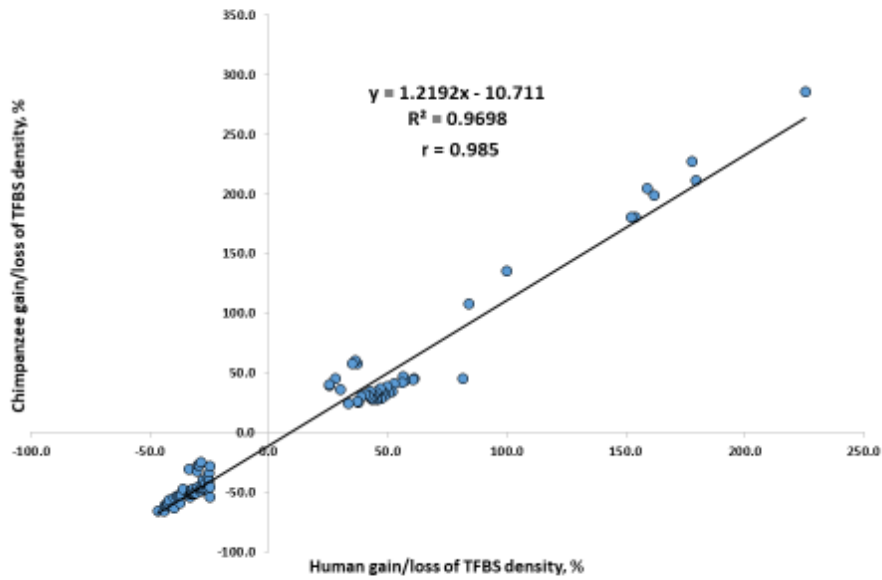

M

**Divergent patterns of TFBS density changes acquired by Chimpanzee and Modern Humans during primate evolution**

Chimpanzee gains: 23 transcription factors

Human gains: 48 transcription factors

Human versus Chimpanzee TFBS density changes, %

**Figure 7.**

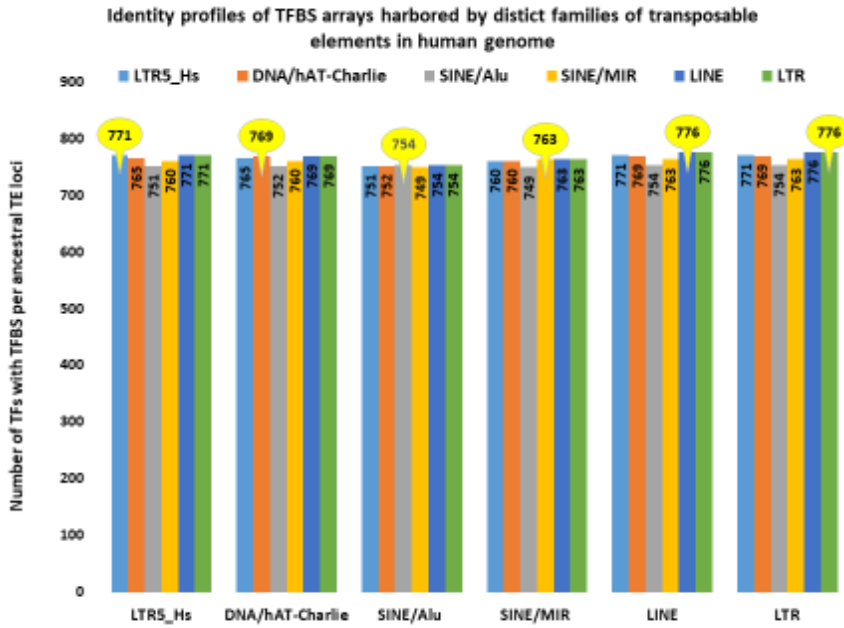

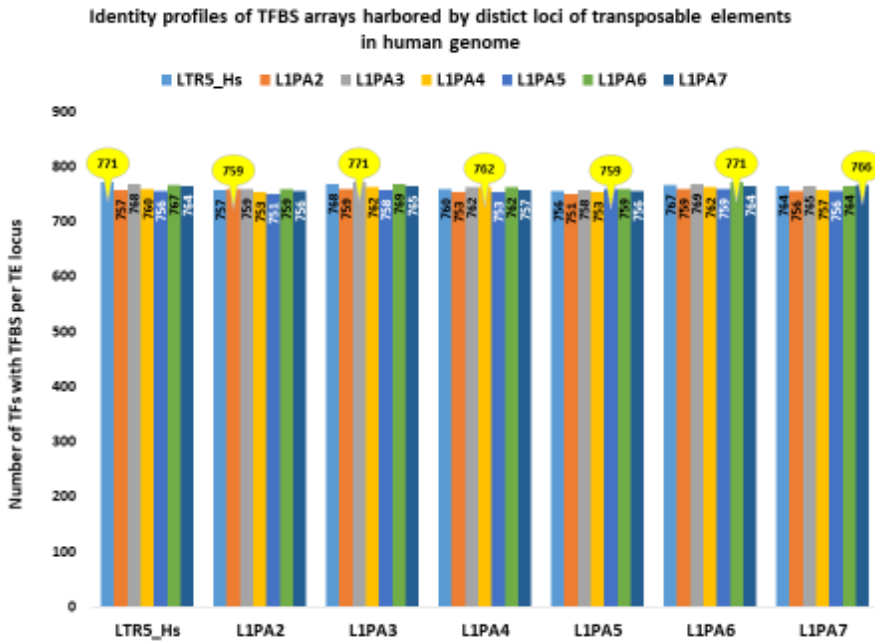
64

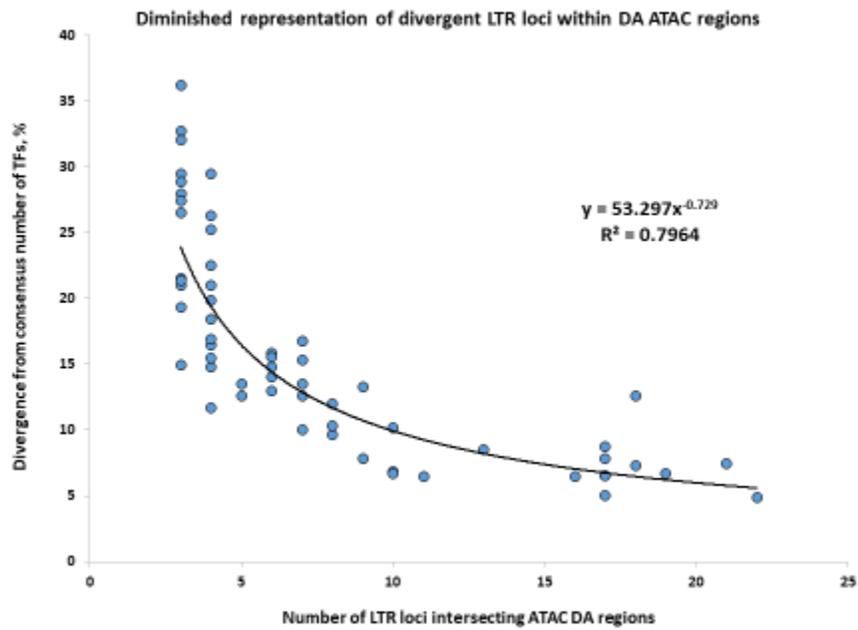

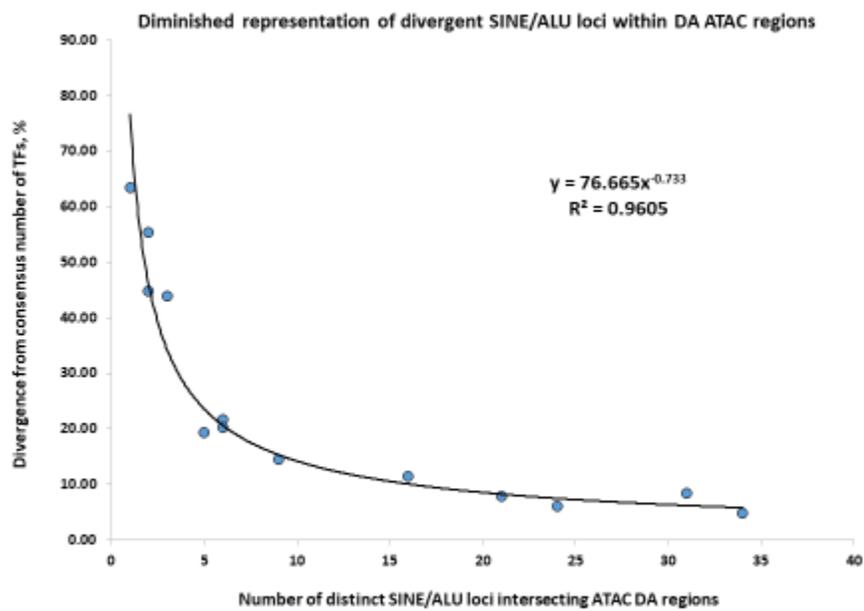

E

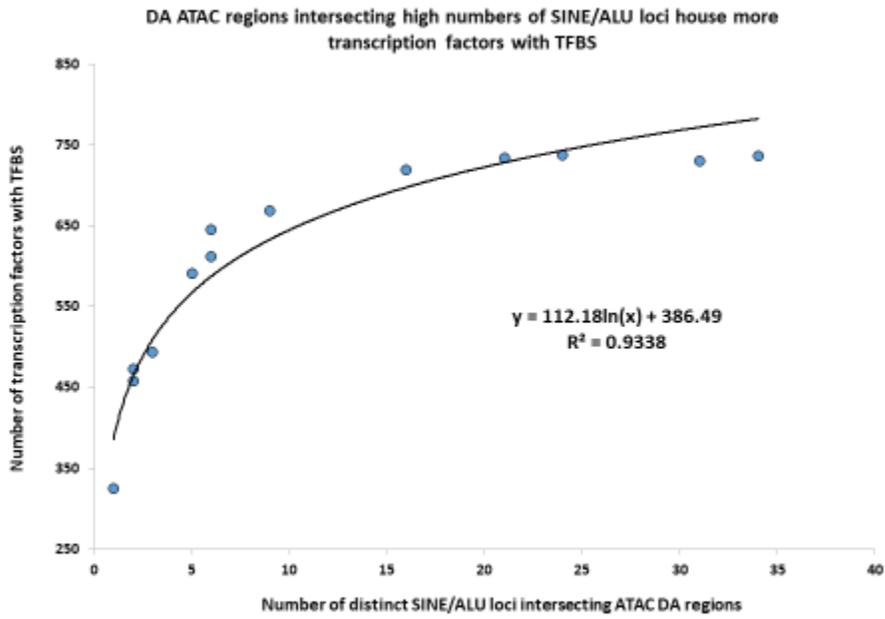

F

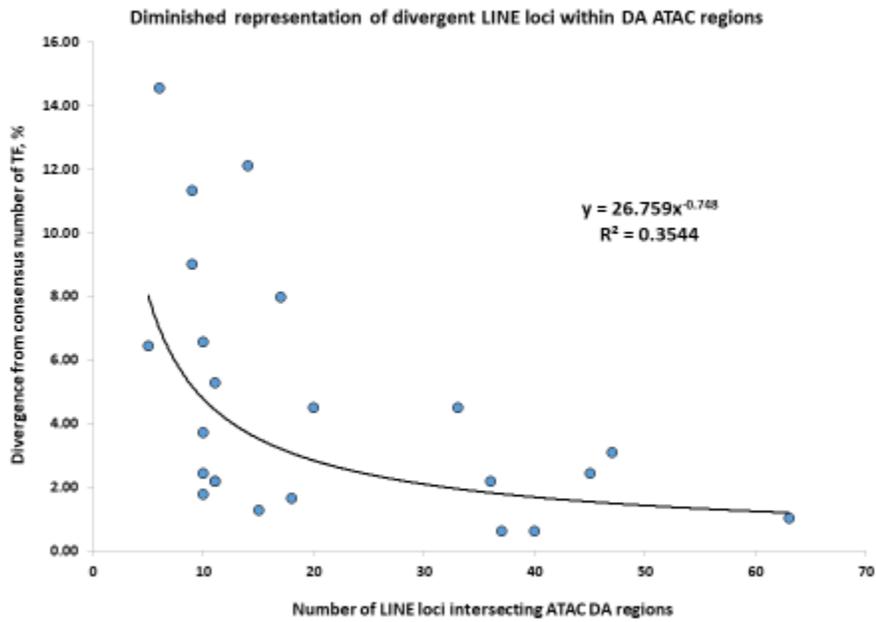

**Figure 8.**

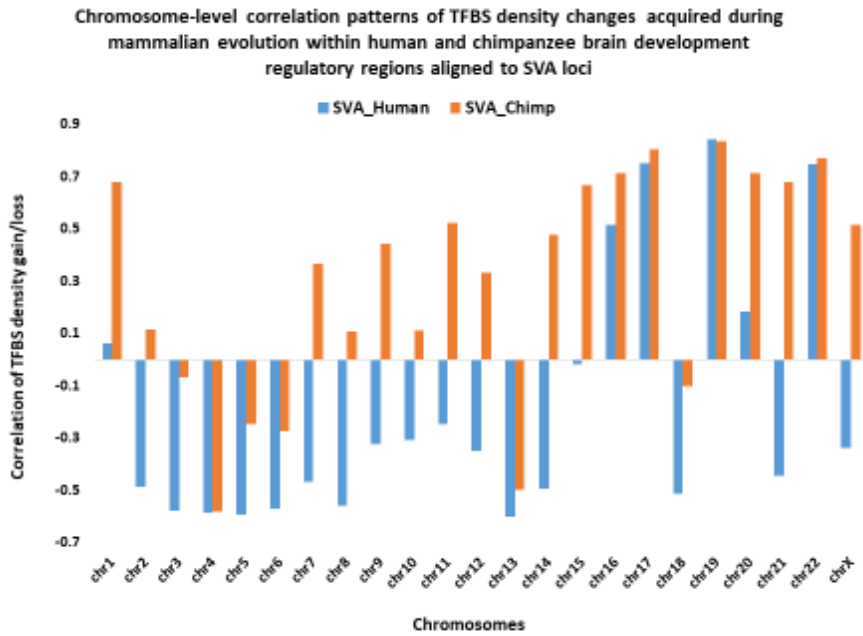

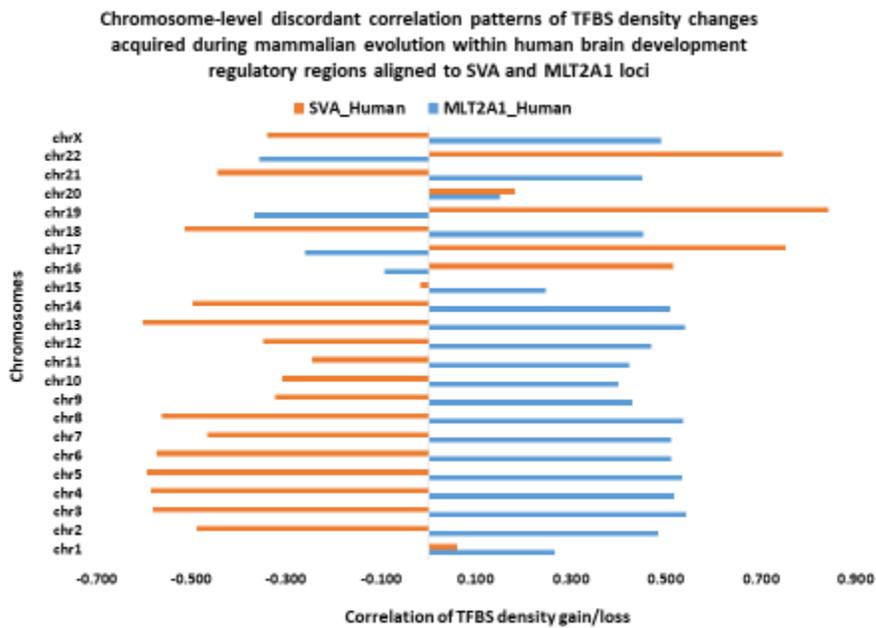

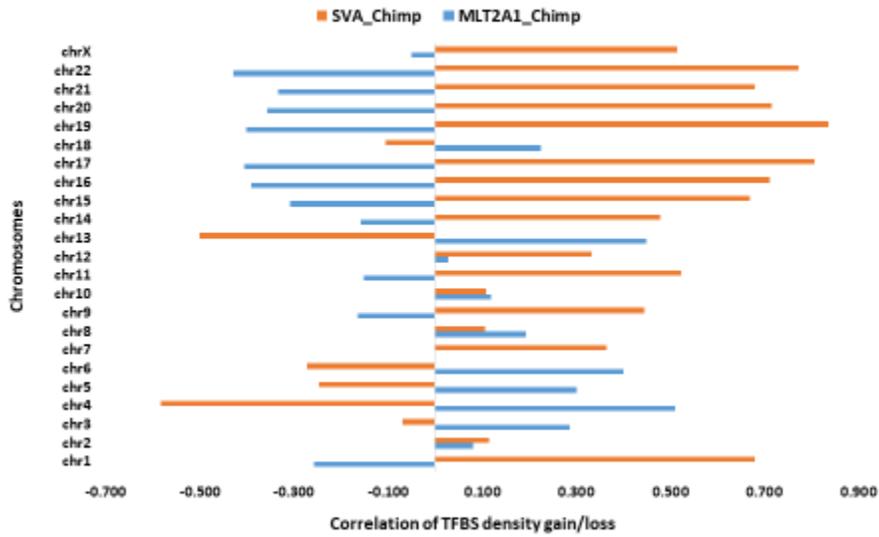

C. Chromosome-level discordant correlation patterns of TFBS density changes acquired during mammalian evolution within chimpanzee brain development regulatory regions aligned to SVA and MLT2A1 loci

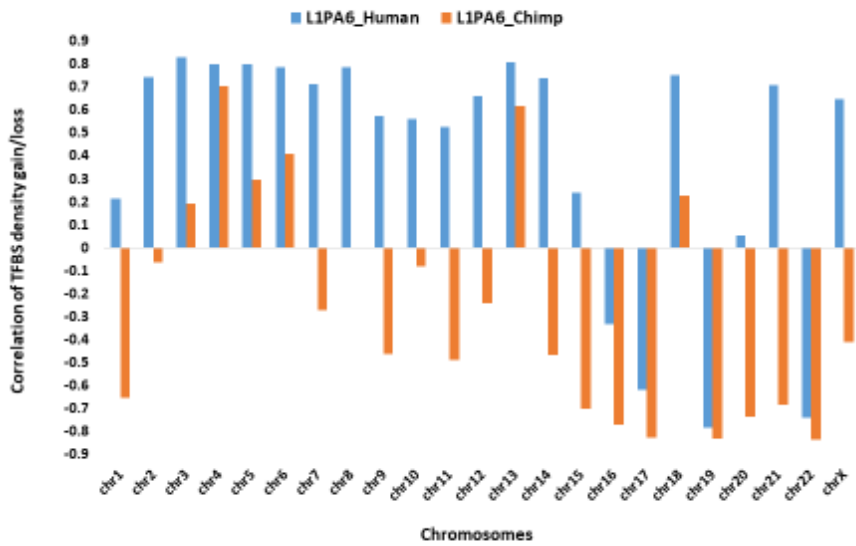

D. Chromosome-level correlation patterns of TFBS density changes acquired during mammalian evolution within human and chimpanzee brain development regulatory regions aligned to L1PA6 loci

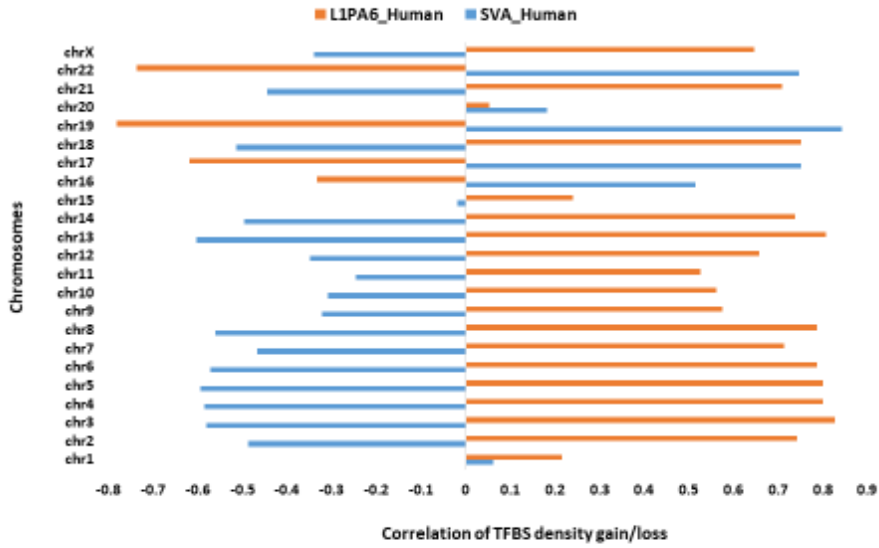

E. Chromosome-level discordant correlation patterns of TFBS density changes acquired during mammalian evolution within human brain development regulatory regions aligned to SVA and L1PA6 loci

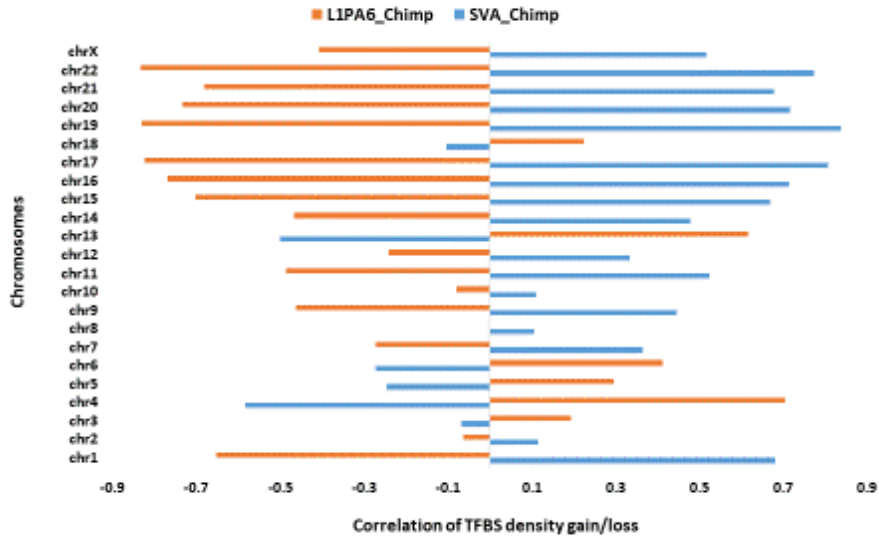

F. Chromosome-level discordant correlation patterns of TFBS density changes acquired during mammalian evolution within chimpanzee brain development regulatory regions aligned to SVA and L1PA6 loci

G

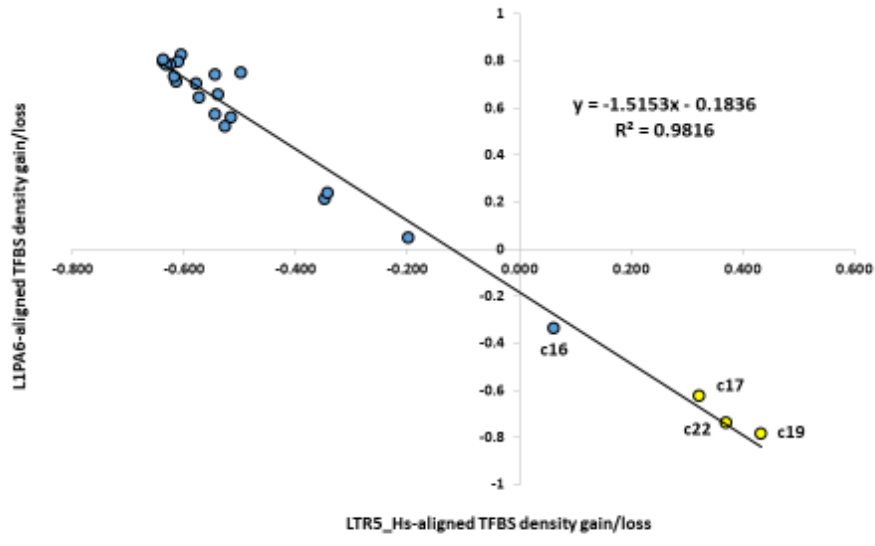

H

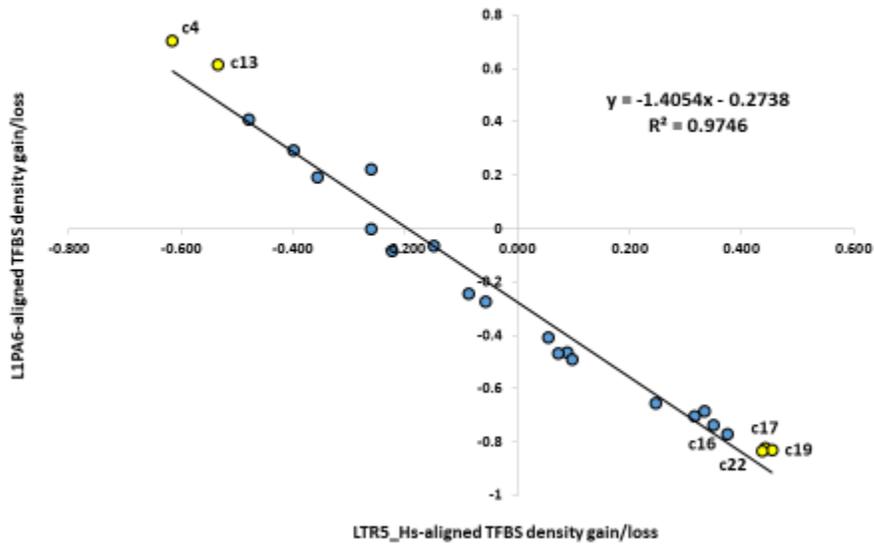

I
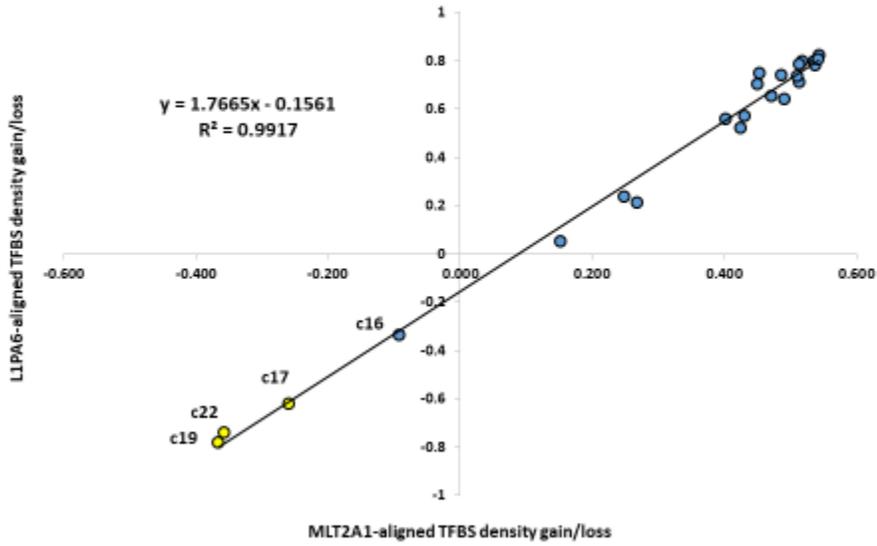

J
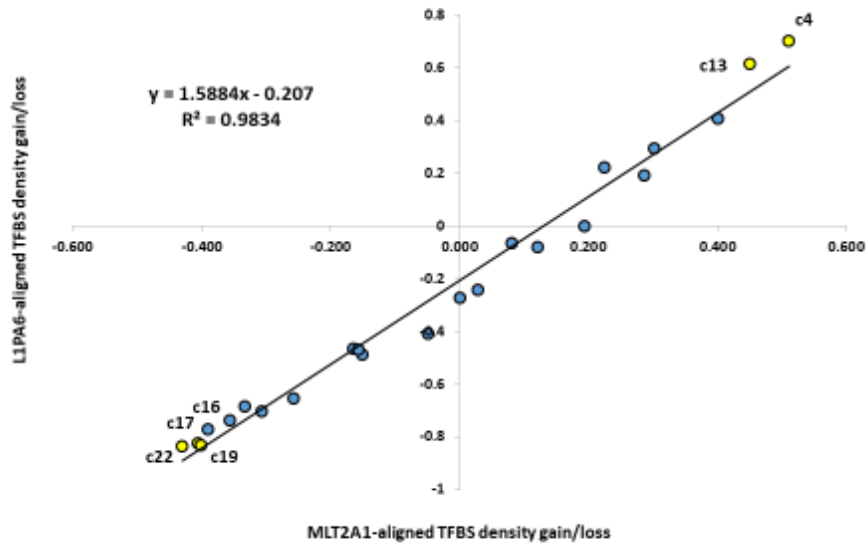

K
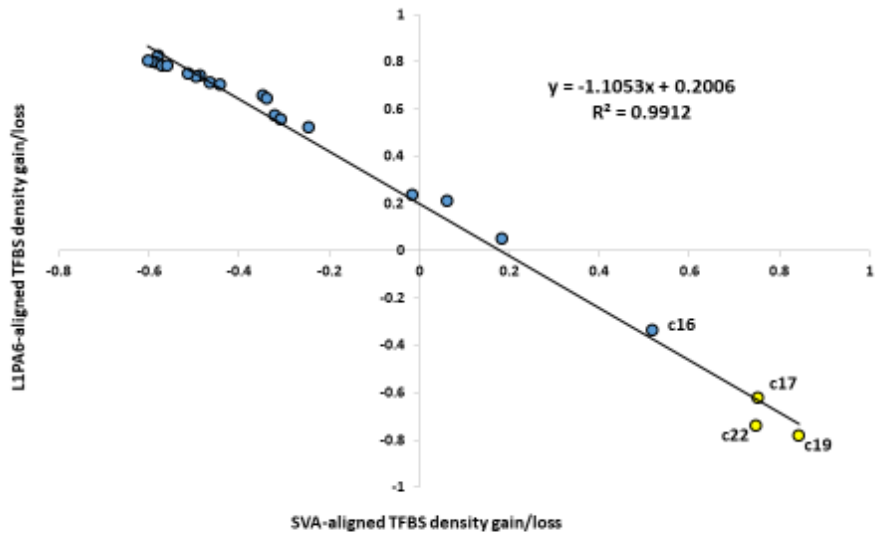

L
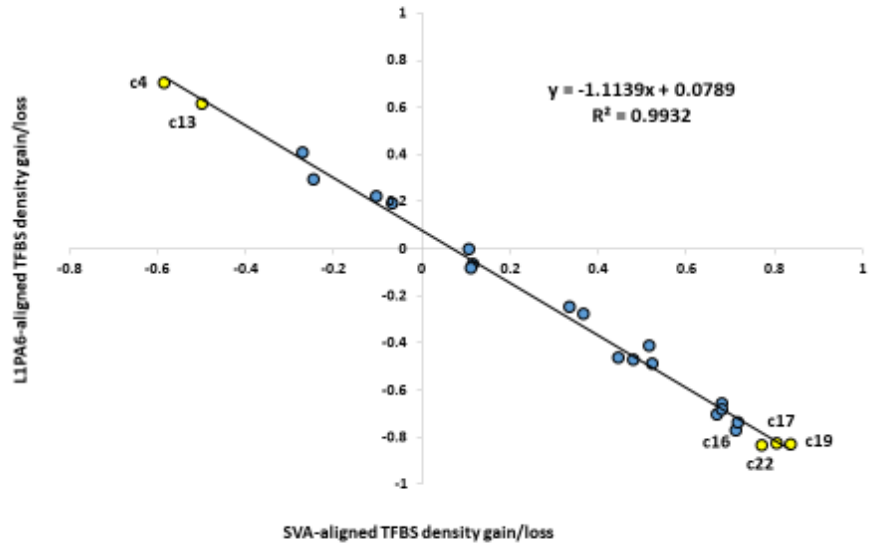

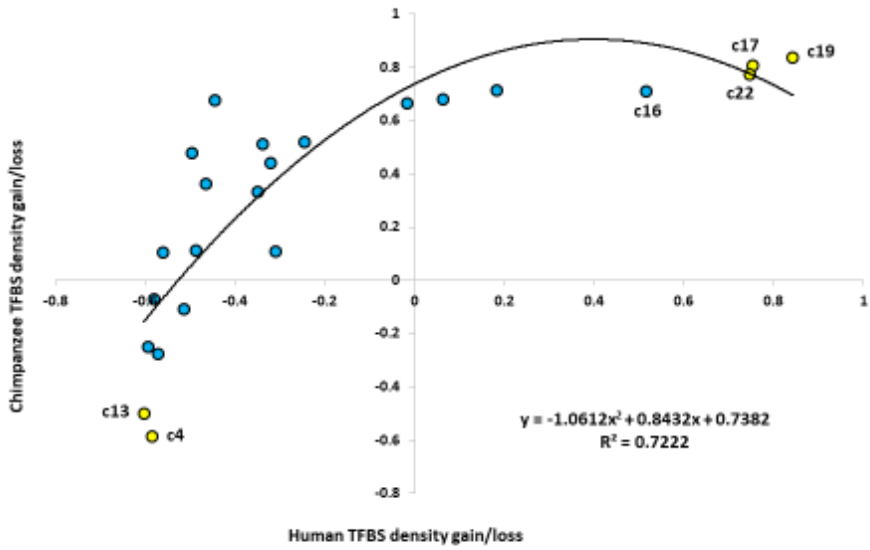

Chromosome-level correlation profiles of TFBS density changes acquired during mammalian evolution within human and chimpanzee brain development regulatory regions aligned to SVA loci

M

$y = -1.0612x^2 + 0.8432x + 0.7382$
$R^2 = 0.7222$

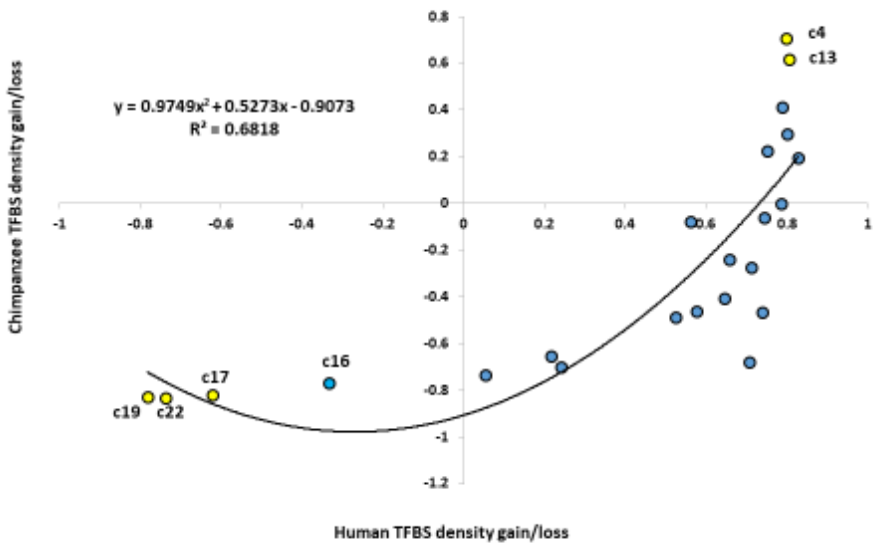

Chromosome-level correlation profiles of TFBS density changes acquired during mammalian evolution within human and chimpanzee brain development regulatory regions aligned to L1PA6 loci

N

$y = 0.9749x^2 + 0.5273x - 0.9073$
$R^2 = 0.6818$

o

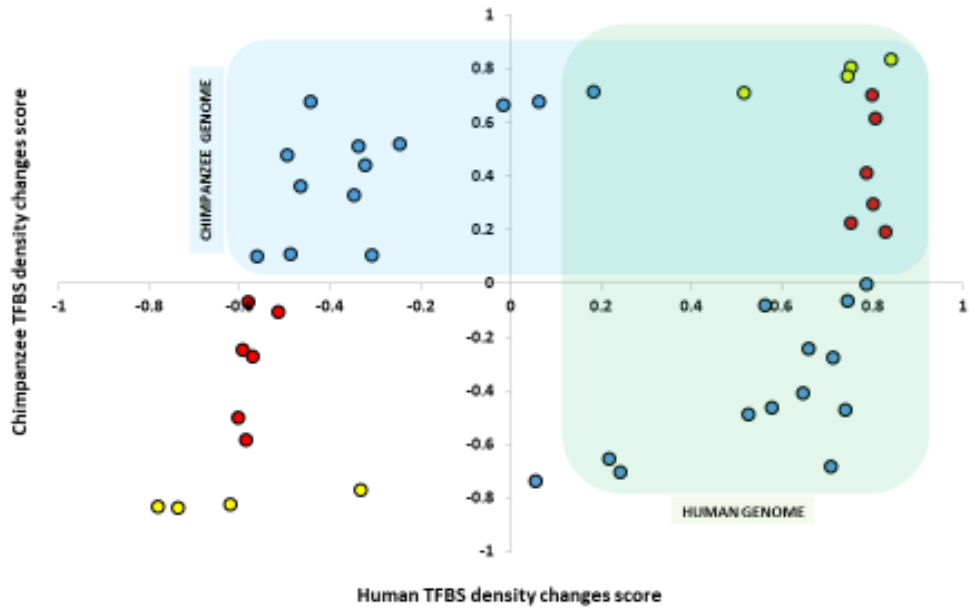

Chromosome-level correlation profiles of TFBS density changes acquired during mammalian evolution within human and chimpanzee brain development regulatory regions aligned to L1PA6 and SVA loci